\documentclass[longauth]{aa}  

\newcounter{daggerfootnote}

\usepackage{graphicx}

\usepackage{txfonts}
\usepackage{natbib}
\bibpunct{(}{)}{;}{a}{}{,}

\usepackage[switch]{lineno}
\usepackage{lscape}             % to rotate a single page table, example in appendix.
\usepackage{xspace}
\usepackage{color}
\usepackage{subfigure}
\usepackage[colorlinks,urlcolor=blue,citecolor=blue,linkcolor=blue]{hyperref} 
\usepackage{newtxtext,newtxmath}

\newcommand{\kms}     {~km~s$^{-1}$\xspace}

\newcommand{\msun}    {~$\mathrm{M_{\sun}}$\xspace}

\newcommand{\mjy}     {~mJy~beam$^{-1}$\xspace}

\newcommand{\malfven}    {$\mathrm{\mathcal{M_A}}$\xspace}

\newcommand{\mfl}   [1]{#1}
\newcommand{\com}   [1]{#1}

\newenvironment{acknowledgments}{%
    \subsection*{Acknowledgments}%
}{}

\begin{document} 
%\linenumbers
%\setlength{\linenumbersep}{10pt}

\titlerunning{MagMaR visits GGD\,27. Magnetic fields overrun by gravity.} 

   \authorrunning{Fern\'andez-L\'opez, M.}
   \title{Magnetic fields in Massive Star-Forming Regions (MagMaR).} 
   \subtitle{VIII. Magnetic field overrun by gravity in GGD~27's accretion streamers}

  \author{  M. Fern\'andez-L\'opez\inst{1,2,3}  
            J. A. L\'opez-V\'azquez\inst{4},
            J. M. Girart\inst{1,5},
            P. Sanhueza\inst{6},
            L. A. Zapata\inst{7},
            P. C. Cort\'es\inst{8,9},
            H. Beuther\inst{10},
            G. Busquet\inst{11,12,5},
            I. W. Stephens\inst{13},
            K. Morii\inst{6,14},             N. A\~nez-L\'opez\inst{15,16},
            C.-F. Lee\inst{4},
            Q. Zhang\inst{14},
            M. T. Beltr\'an\inst{17}, 
            S. Curiel\inst{18},
            F. A. Olguin\inst{19,20},
            E. J. Chung\inst{21},
            P. Saha\inst{4,20}, 
            S. Li\inst{22,23}, 
            P. M. Koch\inst{4},
            J. Hwang\inst{24,25}, 
            C.-Y. Law\inst{17},
            J.-H. Kang\inst{21}
          }

\institute{Institut de Ci\`encies de l’Espai (ICE-CSIC), Campus UAB, Carrer de Can Magrans S/N, E-08193 Cerdanyola del Valles, Catalonia, Spain,
\and Instituto Argentino de Radioastronom\'ia (CCT-La Plata, CONICET; UNLP; CICPBA), C.C. No. 5, 1894, Villa Elisa, Buenos Aires, Argentina,
\and Facultad de Ciencias Astron\'omicas y Geof\'isicas, Universidad Nacional de La Plata, Paseo del Bosque S/N, B1900FWA La Plata, Argentina,
\and Academia Sinica Institute of Astronomy and Astrophysics, No. 1, Sec. 4, Roosevelt Road, Taipei 10617, Taiwan,
\and Institut d'Estudis Espacials de Catalunya (IEEC), Campus del Baix Llobregat - UPC, Esteve Terradas 1, E-08860 Castelldefels, Catalonia, Spain
\and Department of Astronomy, School of Science, The University of Tokyo, 7-3-1 Hongo, Bunkyo, Tokyo 113-0033, Japan,
\and Instituto de Radioastronom\'ia y Astrof\'isica, Universidad Nacional Aut\'onoma de M\'exico, Apartado Postal 3-72, 58089 Morelia, Michoac\'an, M\'exico
\and Joint ALMA Observatory, Alonso de Córdova 3107, Vitacura, Santiago, Chile,
\and  National Radio Astronomy Observatory, 520 Edgemont Road, Charlottesville, VA 22903, USA,
\and Max Planck Institute for Astronomy, Königstuhl 17, 69117 Heidelberg, Germany,
\and Departament de Física Quàntica i Astrofísica (FQA), Universitat de Barcelona (UB), Martí i Franquès 1, 08028 Barcelona, Catalonia, Spain,
\and Institut de Ciències del Cosmos (ICCUB), Universitat de Barcelona, Martí i Franquès 1, 08028 Barcelona, Catalonia, Spain,
\and Department of Earth, Environment and Physics, Worcester State University, Worcester, MA 01602, USA,
\and Center for Astrophysics $|$ Harvard \& Smithsonian, 60 Garden Street, Cambridge, MA 02138, USA,
\and Institut de radioastronomie millimétrique (IRAM), 300 rue de la piscine, 38406 Saint Martin d’Hères, France,
\and Université Paris-Saclay, Université Paris Cité, CEA, CNRS, AIM, 91191 Gif-sur-Yvette, France,
\and INAF-Osservatorio Astrofisico di Arcetri, Largo E. Fermi 5, I-50125 Firenze, Italy,
\and Instituto de Astronom\'{i}a, Universidad Nacional Aut\'onoma de M\'exico (UNAM), Apartado Postal 70-264, DF 04510 M\'exico,
\and Center for Gravitational Physics, Yukawa Institute for Theoretical Physics, Kyoto University, Kitashirakawa Oiwakecho, Sakyo-ku, Kyoto 606-8502, Japan,
\and National Astronomical Observatory of Japan, National Institutes of Natural Sciences, 2-21-1 Osawa, Mitaka, Tokyo 181-8588, Japan,
\and Korea Astronomy and Space Science Institute, 776 Daedeokdae-ro, Yuseong-gu, Daejeon 34055, Republic of Korea,
\and School of Astronomy and Space Science, Nanjing University, 163 Xianlin Avenue, Nanjing 210023, People's Republic of China,
\and Key Laboratory of Modern Astronomy and Astrophysics (Nanjing University), Ministry of Education, Nanjing 210023, People's Republic of China,
\and Institute for Advanced Study, Kyushu University, Japan,
\and Department of Earth and Planetary Sciences, Faculty of Science, Kyushu University, Nishi-ku, Fukuoka 819-0395, Japan.
}

  \abstract
   {\com{Accretion streamers connected to protostellar disks and/or envelopes are thought to transport material across several thousand of au. Whether the motions of the gas comprising these streamers are dominated by gravity, large scale external turbulence or the action of magnetic fields is still under scrutiny.}}
   {\mfl{The aim of this work is to understand} the role of the magnetic fields in the star-formation processes, in particular the role that magnetic forces have in potentially leading flows of gas and the accretion onto the envelopes and disks orbiting protostars.}
   {First, we try to identify the large-scale accretion streamers toward the \com{high-mass Young Stellar Object} GGD~27--MM1 and fit their trajectories using the so-called Mendoza's model, a modification of the classical model of pure gravitational infalling motion of fluid particles in a potential well. Second, we estimate the strength of the magnetic field associated with the streamers. Then, we determine if the streamers are dominated by magnetic or centrifugal forces.}
   {Inspecting the Atacama Large Millimeter/submillimeter Array (ALMA) H$_2$CO cube we were able to identify four accretion streamers spreading up to $\sim$7,000\,au and fit their trajectories in the position-position-velocity space. The polarized continuum emission reveals a good alignment of the magnetic field and the trajectory of the streamers. Using the Davis-Chandrasekhar-Fermi method, we derive estimates for the magnetic field strength, find that the streamers are sub-alfv\'enic, and discuss (after estimating energy terms for turbulence, ordered motions, magnetic forces and gravity) a possible qualitative scenario in which, the gravitational well of the GGD~27--MM1 protostar dominates streamer gas motions over turbulence and magnetic forces at distances of $\sim 3,000$\,au.}
   %{Conclusions}
   {}


   \maketitle
   \nolinenumbers 
%
%-------------------------------------------------------------------

\section{Introduction}
\label{sec:introduction}
Anisotropic accretion through dust and gas streamers has been a recent observational discovery produced thanks to the enormous capabilities of the Atacama Large Millimeter/submillimeter Array interferometer (ALMA) in terms of angular resolution and sensitivity \citep[see review by][and references therein]{2022Pineda}. The new emerging picture on accretion matters \mfl{does not only include} the monolithic collapse of a $\approx10,000$\,au quiescent dense core of dust and gas to form an envelope/disk system. Current evidence suggests that envelopes and/or disks are instead sustained by streamers, coherent flows of gas and dust that transport material from scales of several thousands au to the inner few hundred au \citep[e.g.,][]{2022Lu,2023Olguin,2023Xu,2025Sanhueza,2026Gupta}. In principle, these streamers can replenish the envelope/disk material, producing shocks and heating \com{\citep{2025Liu}}, affecting age indicators \mfl{\citep[such as the bolometric luminosities][]{2023Kuffmeier}}, the final mass, and the formation lifetimes of protostars. Recently, \cite{2025Cortes,2026Huang} have reported sub-alfv\'enic accretion streamers threaded with \mfl{aligned magnetic fields strong enough to remove angular momentum, enabling and guiding} the infall motion of the gas.
Regarding the high-mass protostellar \com{regime}, accretion through streamers may have the property of avoiding the powerful outflow and radiation feedback from the inner parts of these systems \citep{2025Beuther,2025Olguin}. This could allow protostars to grow bigger and for longer periods \citep[e.g.,][]{2006Keto,2010Keto}. 

In addition, channeling material at large distances via streamers could be the path used by different coeval cores and envelopes harboring protostars to compete for the surrounding gas and grow in mass. It has been stated that in crowded environments, the more massive protostars could create stronger potential wells, gaining more mass than their neighbors \citep[e.g.,][]{2001Bonnell}.

This contribution builds upon a previous study reporting accretion through streamers toward the massive protostar GGD~27--MM1 \citep[hereafter MM1, ][]{2023FernandezLopez}. Associated with the IRAS\,18162-2048 source, this source is located at a distance of 1.4~kpc \citep{Anez2020}, and harbors a cluster of Young Stellar Objects (YSOs) detected at different wavelengths \citep{1991Aspin,Gomez1995,Busquet2019}.
The two most massive protostars within this cluster, MM1 and MM2 lie $6\farcs9$ ($\sim9700$\,au) apart from each other, and have been extensively studied \citep[e.g.,][]{Qiu2009,2011FernandezLopez1, 2011FernandezLopez2,2012Carrasco,Girart2018,2023FernandezLopez}. MM1 has a mass of 15-25\msun \citep{Anez2020} and drives the 10~pc long spectacular radio jet associated with the HH\,80-81 objects \citep[e.g.,][]{2010Carrasco,Masque2012,Masque2015,2018Vig,2025RodriguezKamenetzky,2026Fedriani}. The envelope or molecular disk surrounding MM1 (approximately 800\,au in radius) appears to be fed with fresh material from at least three molecular streamers detected in CH$_3$OH, SO$_2$ and H$_2$CO \citep{2023FernandezLopez}. MM2 is likely at an earlier stage of evolution than MM1 \citep{2011FernandezLopez1} and drives a powerful outflow of its own \citep[e.g.,][]{2025LopezVazquez}.

In this \com{study}, we extend the analysis of the streamers \com{potentially} feeding material to MM1's disk/envelope 
%\com{(hereafter we refer to it simply as the disk)} 
through new ALMA observations (Section \ref{sec:observations}). First, we consider a \com{wider perspective of the system, identifying molecular}
streamers on larger scales than \cite{2023FernandezLopez} (up to about 7,000~au). We model the trajectory of these streamers using the formalism explained in \cite{2009Mendoza} (Section \ref{sec:results}), \mfl{as has been done in low-mass star-forming regions by e.g. \cite{2020Pineda,2022ValdiviaMena,2024Gupta,2026Huang}}. Second, we \com{analyze} polarimetric observations related to dust emission that allow us to infer the magnetic field orientation and statistically estimate its strength (also in Section \ref{sec:results}). From the analysis of these results we discuss the distance to the star where the motion of the streaming material is dominated by centrifugal forces rather than magnetic forces (Section \ref{sec:discussion}). Finally, we summarize \com{our findings} in Section \ref{sec:conclusions}.

\begin{figure*}[t!]
\centering
\includegraphics[scale=0.525]{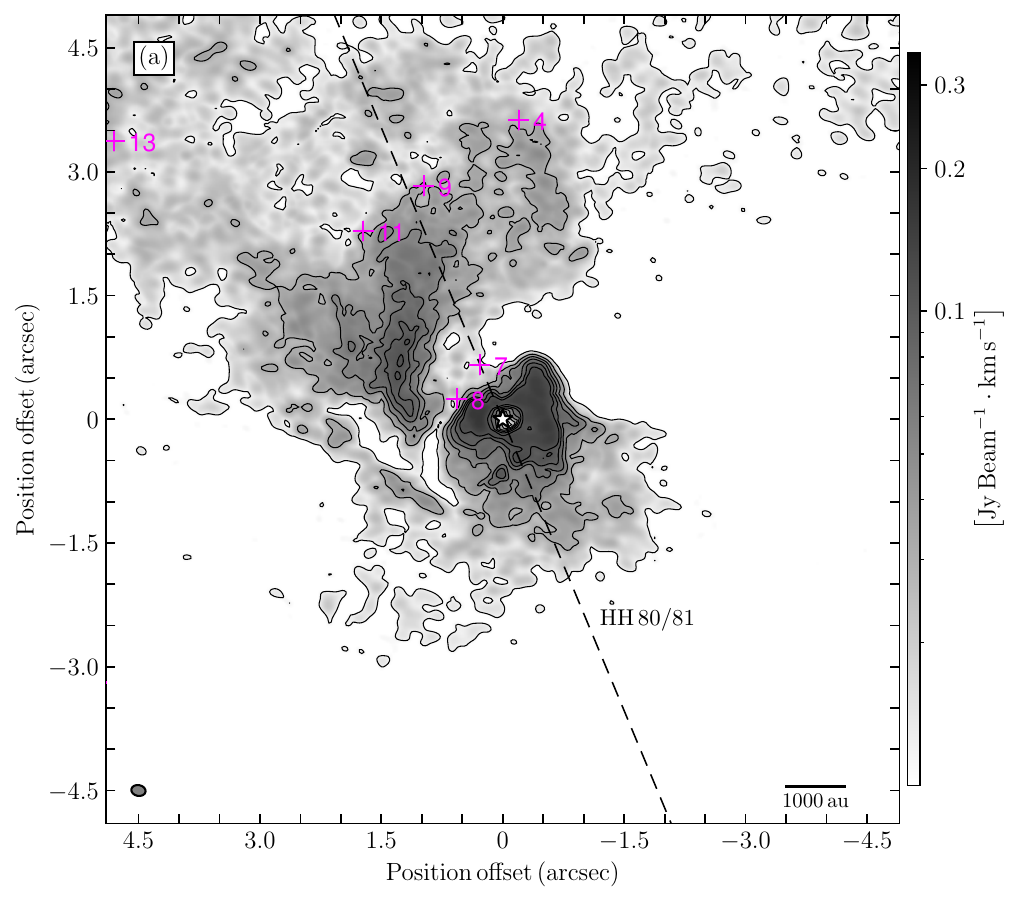}\includegraphics[scale=0.525]{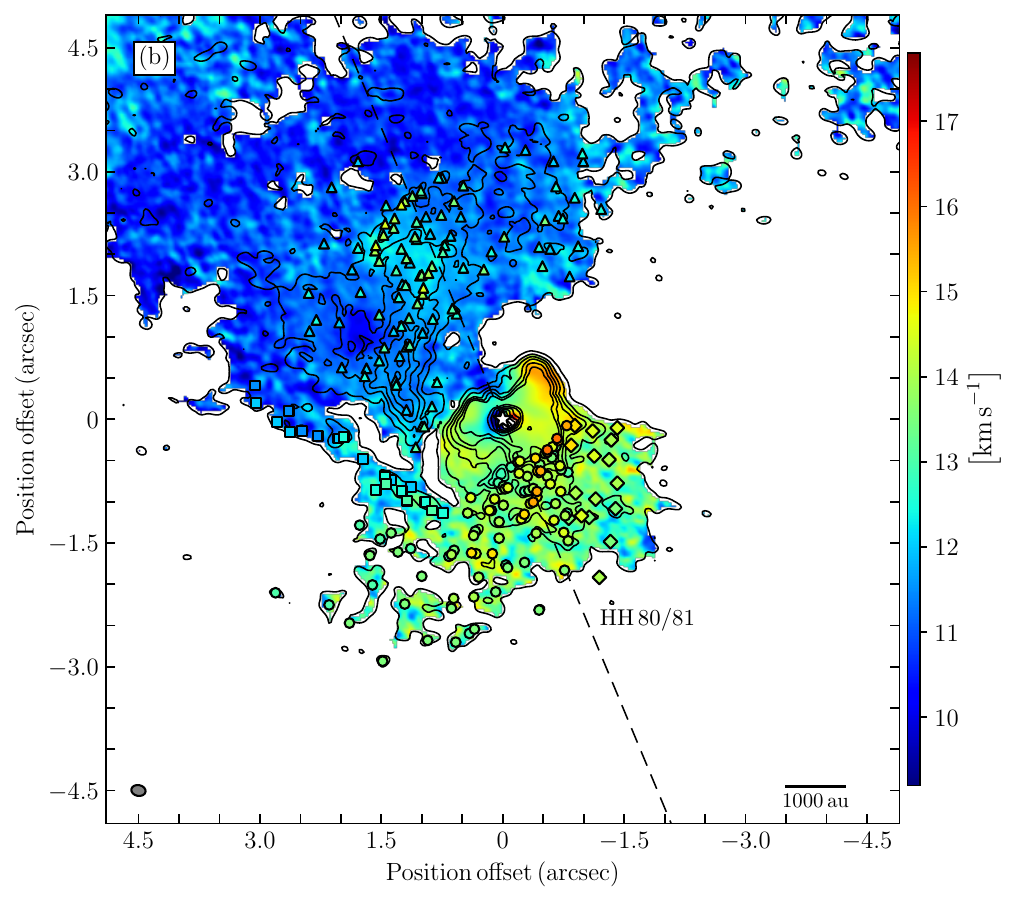}
\begin{flushleft}
\includegraphics[scale=0.525]{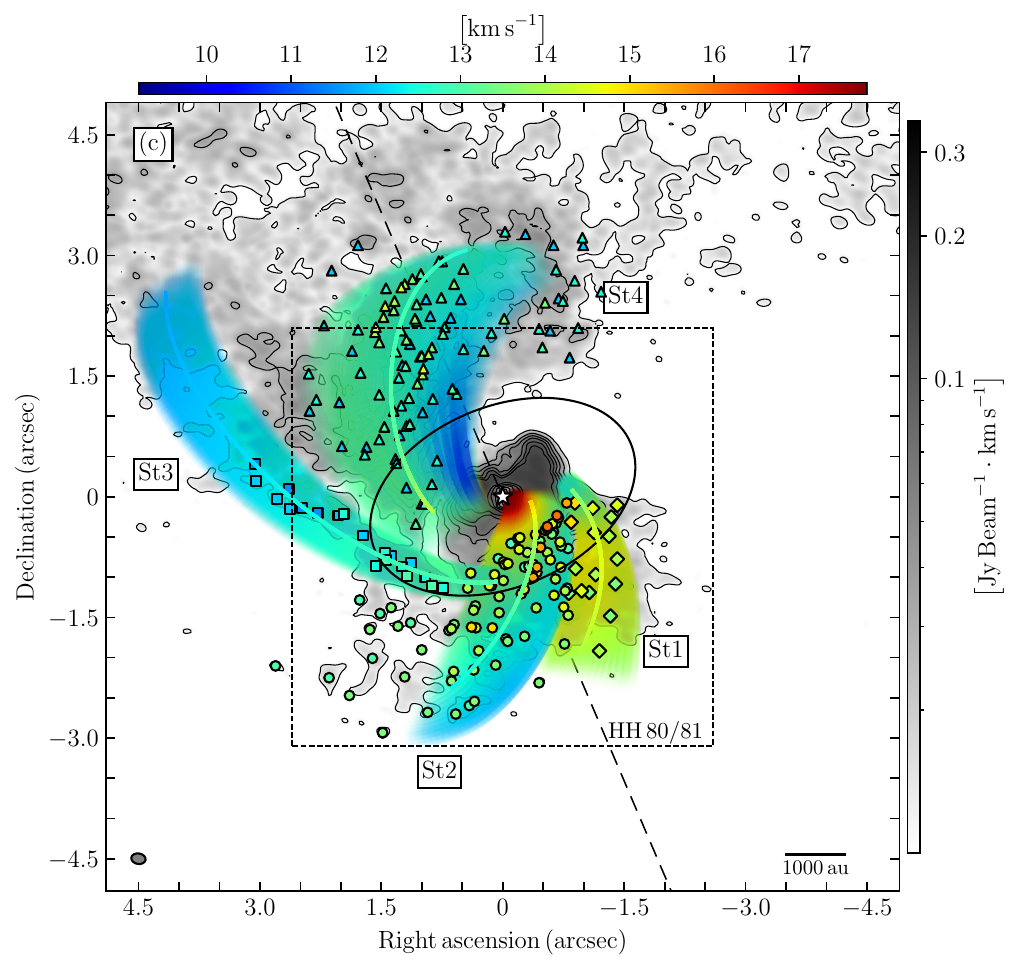}\includegraphics[scale=0.525]{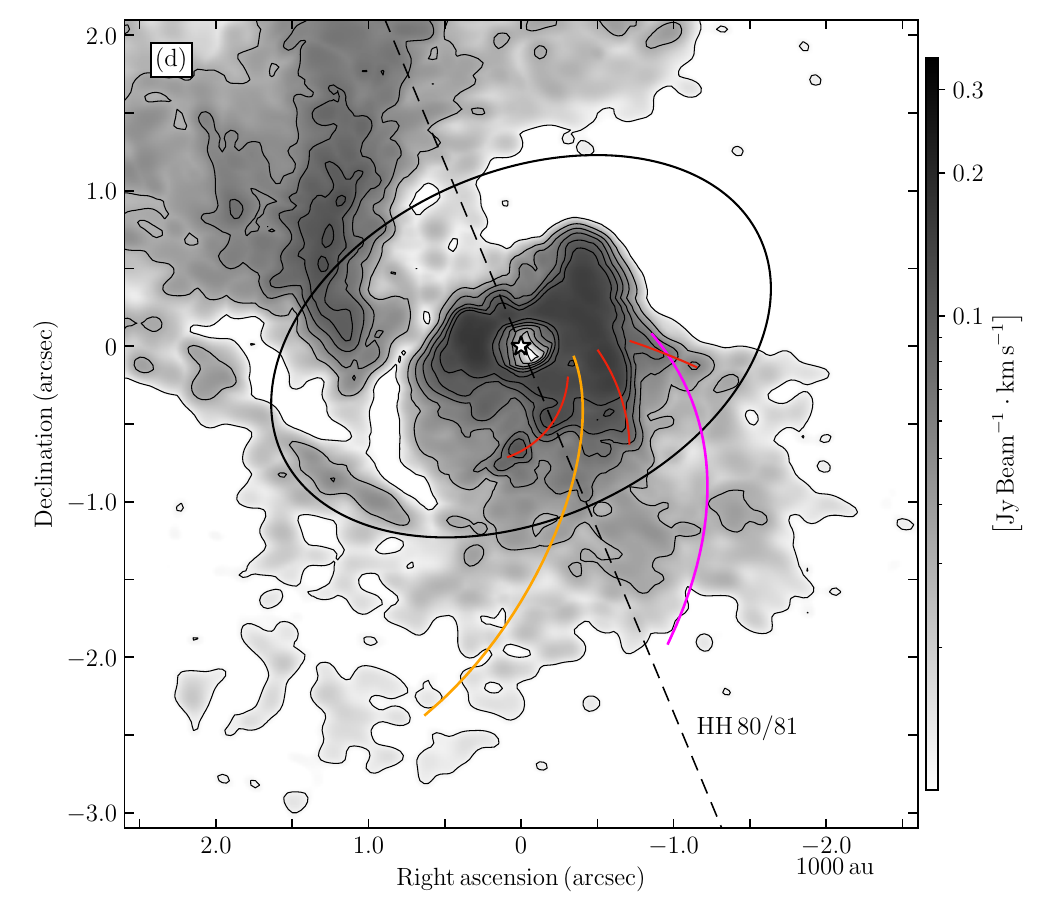}
\end{flushleft}
\caption{\com{GGD 27-MM1 molecular line emission of H$_2$CO. (a) ALMA moment 8 or peak intensity map. (b) ALMA first moment or intensity-weighted velocity map (color scale) overlaid with moment 8 contours. (c) ALMA moment 8 map overlaid with color-shaded areas representing the line-of-sight velocities of four streamers following a funnel-like structure of material infalling toward the central protostar. The solid central line indicates the best-fit model, whose parameters are listed in Table \ref{tab:mendozaparameters}. \mfl{(d) Zoom-in of the moment 8 map shown in panel (a).} Magenta crosses in panel (a) represents the position of the continuum sources reported by \citet{Busquet2019}. In panels (b) and (c), the diamonds, circles, squares, and triangles mark the positions of the condensations identified in the velocity cube (see text), with the color indicating their line-of-sight velocity. \mfl{The dashed square in panel (c) marks the zoomed-in area of panel (d). Red lines in panel (d) mark three dense ridges showing possible streamer connections, whereas the pink and orange lines show the spine of the streamers St1 and St2, respectively.} The contour levels start at 3$\sigma$ and increase in steps of 5$\sigma$ up to 28$\sigma$, where $\sigma= 4.11$ \mjy \kms. The dashed black line indicates the direction of the protostellar jet HH~80/81. The synthesized beam is shown in the bottom-left corner.}}
\label{fig:mom8streamers}
\end{figure*}

\begin{figure*}[t!]
\centering
\includegraphics[width=\linewidth]{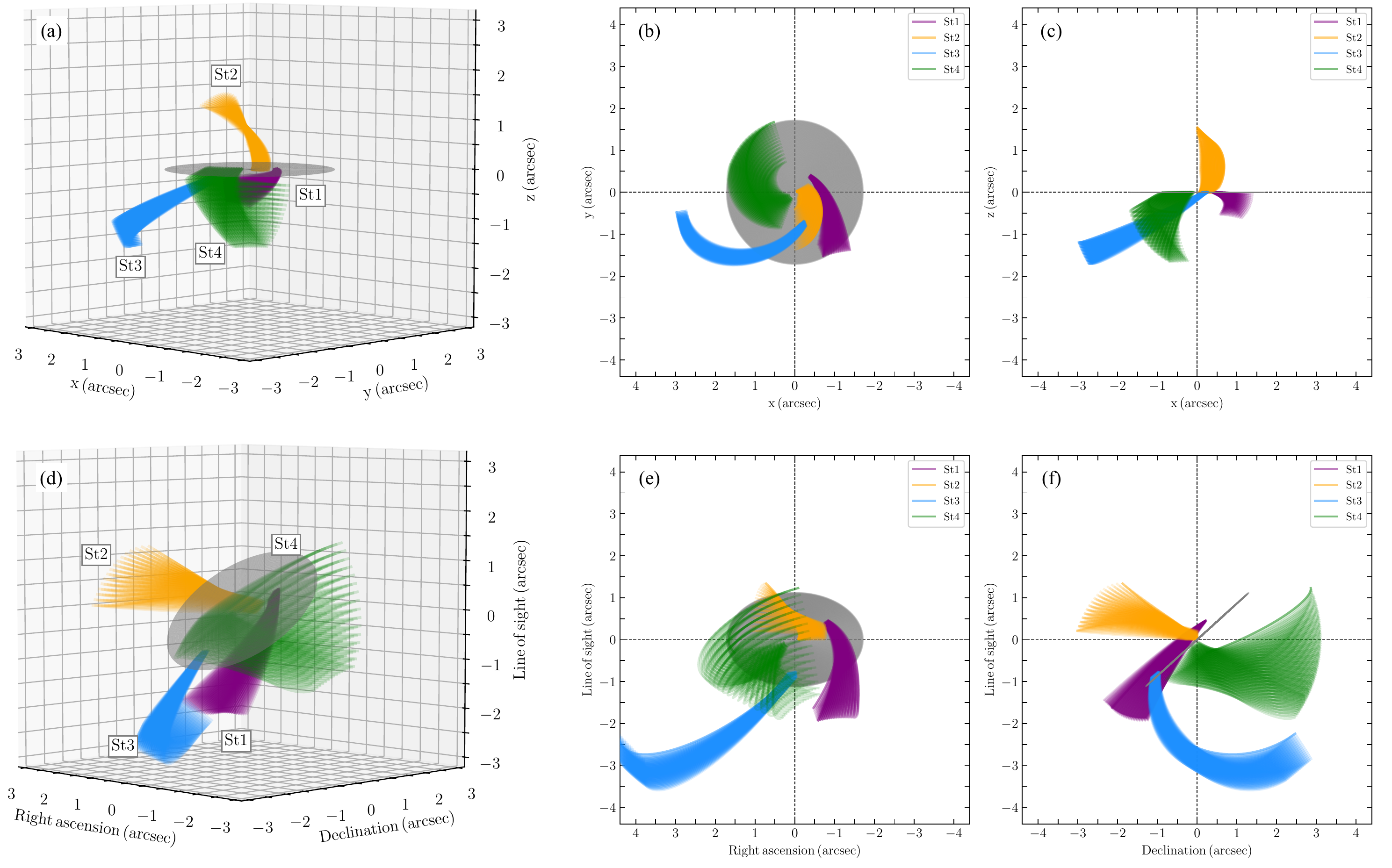}
\caption{Best-fit models for the four accretion streamers. (a) Three-dimensional diagram in the reference frame of the central protostar. (b) Projections of the four streamers in the $xy$ plane. (c) Projections of the four streamers in the $xz$ plane. (d) Three-dimensional diagram in the reference frame of the observer, where the $z-$axis represents the line of sight. (e) Projections of the four streamers in the right ascension-line of sight plane. (f) Projections of the four streamers in the declination-line of sight plane. The gray disk corresponding to the Ulrich's radius (see the text).}
\label{fig:projections}
\end{figure*}

\section{Observations}
\label{sec:observations}
This contribution uses two different ALMA datasets (projects 2015.1.00480.S and 2017.1.00101.S; PIs: Josep Miquel Girart and Patricio Sanhueza, respectively). Each is used with a different purpose within this work: the former to image the molecular emission of the gas streamers infalling toward MM1 at large scales; the latter to map the continuum emission and trace the magnetic field morphology on these structures. Both datasets provide with a new view of the large scales ($\approx$thousands of au) of the GGD~27 system.  

The 2015.1.00480.S \com{dataset, previously presented in \cite{2023FernandezLopez} focusing on the immediate surroundings of the MM1 \mfl{disk/envelope}, includes observations of the H$_2$CO ($4_{1,3}-3_{1,2}$) line emission (rest frequency 300.836635\,GHz and upper energy level of 47.9\,K) sampled with a spectral resolution of 0.5~\kms. The velocity cube has an rms noise level of 2.1\mjy and a synthesized beam of $0\farcs18\times0\farcs14$ with PA of $80^{\circ}$. The maximum recoverable angular scale of these Band~7 data is about $1\farcs2$. In the present work we revisit these data to explore molecular gas structures on larger spatial scales.}

The project 2017.1.00101.S data \citep[for references on the \com{Magnetic fields in Massive star-forming Regions,} MagMaR project, see e.g., ][]{2021FernandezLopez,2021Cortes,2021Sanhueza,2024Cortes,2024Saha,2024Zapata,2025Sanhueza,2026Hwang} contain the polarized dust emission at 1.2~mm (ALMA Band~6). A study of the polarized emission toward the disk of MM1 has already been reported \com{by \cite{2018Girart}. In contrast, the observations presented in this contribution focus on larger spatial scales and avoid discussing the properties of the disk.}
The 2017.1.00101.S data were taken on 2018 September 25 with the 12~m array, and have a total effective line-free bandwidth of 6.8~GHz \citep[see ][for the line removal procedure]{2021Olguin}. The observations were calibrated by performing a standard procedure, which was followed by \com{three iterations of phase-only self-calibration with the last one using a solution interval of 10~seconds to retrieve} naturally weighted Stokes I, Q, and U images. Using the natural weighting in the deconvolution process provides images with slightly lower noise and therefore, more sensitivity. The final Stokes~I image has a rms noise level of 0.12\mjy and a synthesized beam of $0\farcs34\times0\farcs27$ with a P.A. of $76\degr$.  The maximum recoverable angular scale of these data is about $3\farcs0$. For the Stokes~Q and U images the rms noise level is 0.032\mjy and 0.034\mjy, respectively. From these, we prepared images of the debiased linear polarized intensity \citep{2006Vaillancourt} and the magnetic field orientation, after applying a $90\degr$ rotation to the electric half-vector \com{polarization} position angles.  

\com{These interferometric data are not combined with single-dish and therefore some extended emission may be filtered out. However, since the maximum recoverable scale of the datasets is of the order of the width of the structures analyzed we do not expect to miss a huge amount of flux from them. In fact, the 2017.1.00101.S Band~6 continuum image recovers practically all the flux density (we measured 1.415 Jy at 250~GHz) expected from the much lower angular resolution Submillimeter Array 220~GHz observations (1.2 Jy at 220~GHz, measured with a beam size $8\farcs1\times3\farcs0$ and maximum recoverable scale of $22\arcsec$, exceeding the size of the continuum emission of the whole region) reported in \cite{2011FernandezLopez1}.  Taking into account an average positive spectral index for the dust, we estimate a potential flux loss $\lesssim15$\%, which could also be accounted by the absolute flux calibration uncertainties (expected to be about $10$\% and $20$\% for ALMA and SMA, respectively).}

\section{Results and analysis}
\label{sec:results}
\com{This work extends the analysis of the streamers that are infalling} toward the MM1 envelope/disk \citep{2023FernandezLopez}. \com{While the earlier study concentrated on the central part of the system, where streamers interact with the \mfl{disk/envelope system}, this study broadens the scope, tracing the streamers up to $\sim7,000$\,au upstream.} The following sections \com{present} (i) a kinematic and morphological fitting to the large-scale H$_2$CO data \com{using the infalling streamer model proposed by} \citep{2009Mendoza}, (ii) an estimate of the density, mass and momentum of the streamers, and (iii) an analysis of the polarized dust emission tracing the magnetic field of the streamers which is used to statistically estimate the magnetic field strength using the Davis-Chandrasekhar-Fermi, DCF, approximation \citep{1951Davis,1953Chandrasekhar}.

\subsection{Streamer trajectory fitting}
\label{subsec:streamers}

A moment 8 map (peak intensity) and \com{a moment 1 (intensity-weighted mean velocity)} map of the large-scale emission of the H$_2$CO molecular line associated with MM1 is presented in Figure \ref{fig:mom8streamers}, where we mark four curved streamer--like structures that connect the outer molecular cloud with the accretion disk and envelope previously reported by \citet{2023FernandezLopez}. These structures are named St1, St2, St3, and St4, starting in the southwest with St1 and continuing clockwise. 
The streamers are better seen in the channel velocity map where they have been identified (see Figure \ref{fig:channelsmaps} in Appendix \ref{app:channel}). This Figure shows that these structures are coherent in velocity with spatial displacements that can be regarded as rotation plus infalling motions toward MM1's \mfl{disk/envelope}. While the streamers display a spiral pattern all twisting in the same direction and could be even seen as a single rotating entity, we decide to follow the overdensities of the gas and dust in the form of elongated ridges and interpret them as possible streamers. The boundary between the St1 and St2 streamers is related \mfl{to the association of St2 to the inner streamer S2} reported in \cite{2023FernandezLopez} \mfl{and the presence of three small ridges (marked by red lines) coming out south of the envelope (see panel (d) in Figure \ref{fig:mom8streamers}) with different orientations. While we associate the two eastern ridges (corresponding to the inner S2 and S3 streamers) to St2, we associate the westernmost ridge to St1.} The streamers St3 and St4, identified here for the first time, seem more well spatially separated. The streamer St4 could be more complex than just one entity.

The St1 streamer (magenta curve and markers in Figure \ref{fig:channelsmaps}) runs from southwest of the disk to its northwest edge, and it is seen at velocities from 13.5\kms to 16\kms (note that the system velocity is 12.1\kms); \mfl{it corresponds with the inner westernmost ridge seen in Figure \ref{fig:mom8streamers}}. The St2 streamer (associated with the inner \mfl{S2 and maybe S3 streamers}) goes from southeast to south of the MM1 position at velocities from 13.0\kms to 16.0\kms (yellow curve). \mfl{St2 shows a slightly smaller average velocity (13.9\kms) than St1 (14.3\kms).} The St3 streamer is clearly seen at velocities between 12.0\kms and 13.0\kms running from the east and \mfl{likely} connecting the \mfl{envelope} toward its southeast edge (blue curve). Finally the St4 streamer (green curve and markers) has a wider funnel-like aspect, perhaps hinting a bifurcation, with a well-defined north-south ridge seen at velocities between 12.0\kms and 13.0\kms, and connecting the \mfl{envelope} at its  northeast edge.
For each of these curved structures, we \com{manually} selected gas condensations on the H$_2$CO channel maps (Figure \ref{fig:channelsmaps}). Condensations are \com{identified as independent peaks of emission channel by channel. They have typical sizes of a beam and are included in a streamer list} when their intensity is at least 5$\sigma$. \com{We avoid to include condensations close to or within the MM1 \mfl{disk/envelope system}, where the angular resolution is not enough to distinguish their path and some of the streamers seem to merge.} The locations of these condensations are plotted in Figure \ref{fig:mom8streamers} as stars, circles, squares, and triangles for St1, St2, St3, and St4, respectively. 

Considering the position and the velocity of the condensations, we fitted the accretion streamer model following the formulation presented by \citet{2009Mendoza} . This model considers that the material originally lies in a rotating molecular cloud undergoing gravitational collapse. The cloud has a defined size, and the particles begin their collapse with an initial velocity in the radial direction (i.e., toward the protostar), following an elliptical trajectory in which the gas conserves energy and angular momentum. In this idealized scenario, pressure effects \com{(including magnetic field effects)} are assumed to be negligible. Mendoza's model is an extension of the Ulrich's model \citep{1976Ulrich}, where the particle's initial velocity in the radial velocity is zero. 

The trajectory of the gas is described by four free parameters: $\mu$, which defines the ratio of a fixed radius in the Ulrich's model, corresponding to the centrifugal radius of the cloud, to the radius of the initial cloud; $\nu$, which is the ratio between the initial radial velocity and the Keplerian velocity at the aforementioned Ulrich's radius. The other two free parameters, $\theta_0$ and $\phi_0$, are related to the initial polar and azimuthal positions from which the particle starts its collapse toward the central star.  
The parameters $\mu$ and $\nu$ depend on the rotational and infall properties of the cloud: $\mu$ is linked to the rotational motion through the Ulrich's radius and thus to the initial angular momentum, while $\nu$ is related to the infall motion and together reflect the initial energy of the cloud, including gravitational, kinetic, and rotational contributions. 

To explain the trajectory and the observed velocity, we adopted a stellar mass of $M_*=20$ \msun, an inclination angle of the system with respect to the plane of the sky of $42^{\circ}$  (\citealt{Anez2020}), and an Ulrich's radius of 2400 au (at this radius, the rotational velocity of the streamers becomes the Keplerian velocity). To obtain the best parameters of the model, we use the Asexual Genetic Algorithm (AGA) developed by \citet{Canto2009}, \com{which is a computationally cheap and accurate method to find the minimum of a given function.} The results are shown in Table \ref{tab:mendozaparameters}. From the best-fit solutions, we derive the eccentricities ($e$) of the streamers, which are also reported in Table \ref{tab:mendozaparameters}. Since the parameters $\mu$ and $\nu$ include both the initial rotational and infall properties of the cloud, the large variations observed in these parameters among the streamers do not necessarily translate into significant differences in their eccentricities. Given that the obtained parameters allow us to reproduce the spine of the streamer, we consider a bundle of trajectories whose initial positions ($\theta_0$, $\phi_0$) surround the position of the fitted spine trajectory, forming a patch determined by variations in the initial angles, $\Delta \theta_0$ in the polar direction and $\Delta \phi_0$ in azimuthal direction. These variations, which also encompass the statistical uncertainties associated with the parameter estimation, are shown in Table \ref{tab:mendozaparameters}. The trajectories of this bundle of streamers, create a tube-like structure of material following elliptical trajectories. These trajectories are presented with their corresponding observed line-of-sight velocity in Figure \ref{fig:mom8streamers}a. \com{If governed by gravity, the modeled trajectories predict that the} furthest particles of the St1 , St2, St3 and St4 streamers will spend 5300~yr, 5500~yr, 29000~yr, and 12300~yr to reach the molecular \mfl{disk/envelope} of MM1 \citep[about 790~au from the central star][]{2023FernandezLopez}, respectively. \com{These values were derived by numerically integrating in time the expression for the radial position of the particles at the end of each streamer. We will use them as the lifetimes of the four streamers.} 

Figure \ref{fig:projections} presents \com{models for} the four streamers associated with the protostellar system MM1 in the reference frames of the central protostar (upper panels) and the observer (bottom panels). The three-dimensional paths of the material are shown in Figures \ref{fig:projections}a and d, while the projections onto the $xy$ and $xz$ planes are presented in \ref{fig:projections}b and c, respectively, and the projections along the line of sight, in right ascension and declination, in Figures \ref{fig:projections}e and f, respectively. The line-of-sight positions are reconstructed numerically using the three-dimensional Mendoza model, constrained by the observed projected positions ($x^{\prime}$, right ascension, and $y^{\prime}$, declination), the line-of-sight velocities, and the inclination angle of the system. The reconstructed $z$ and $z^{\prime}$ coordinates are then used to estimate the best-fit model parameters ($\theta_0$, $\phi_0$, $\mu$, and $\nu$), allowing the model to simultaneously reproduce the observed morphology and kinematics of the streamers.

\begin{figure*}[t!]
\centering
\includegraphics[width=\linewidth]{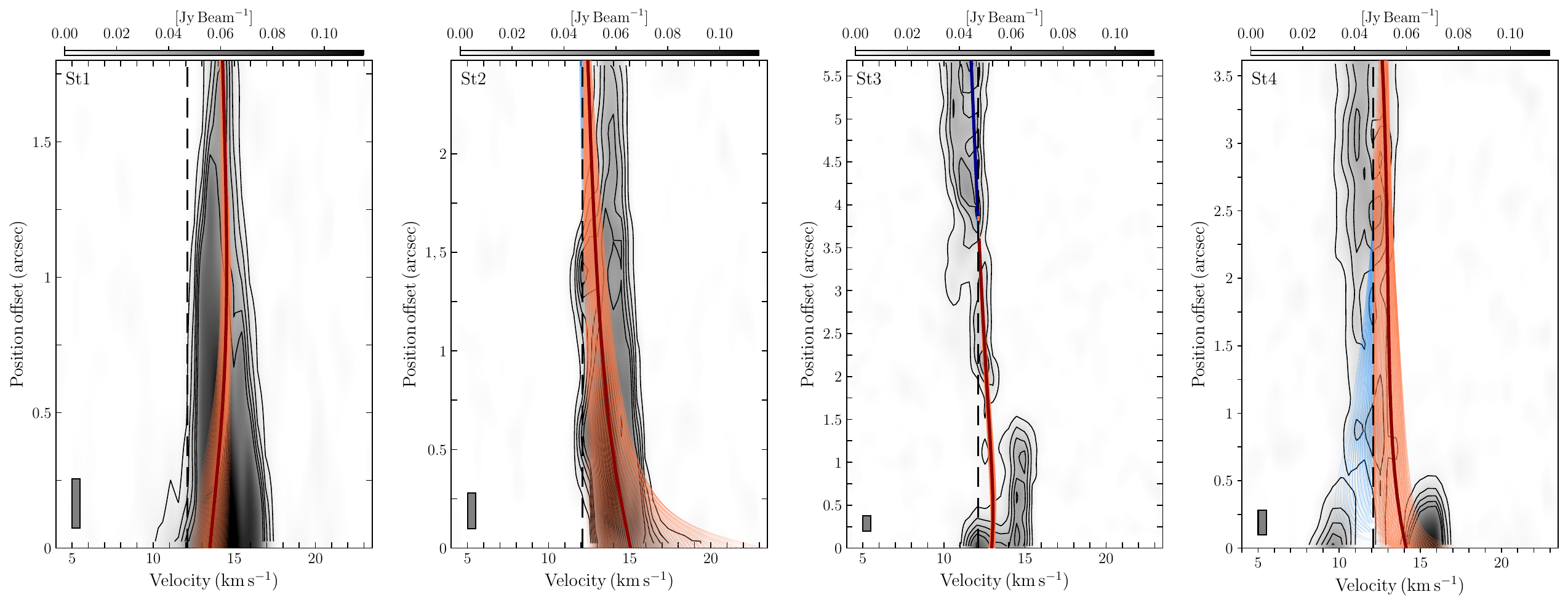}
\caption{Position-velocity diagrams along the streamers St1, St2, St3, and St4 for our best-fit models. \com{Black contours and gray scale color represent the brightness intensity. The colored lines correspond to our model results: thick lines show the best-fit trajectories of the spine streamers, and thin lines indicate the fitted trajectories of the bundle streamers. The color of each line reflects the line-of-sight velocity: blue for blueshifted and red for redshifted emission. The black dashed line in all panels marks the systemic velocity $V_{sys} = 12.1$\kms. The gray bars in the lower-left corner indicate the channel width and the angular resolution.}
}
\label{fig:pvdiagrams}
\end{figure*}

To compare the observed \com{and modeled streamer velocities} in more detail, four position-velocity (PV) diagrams along the streamers are depicted in Figure \ref{fig:pvdiagrams}. The PV diagrams were extracted following the central lines shown in Figure \ref{fig:mom8streamers}c, using a width equal to one beam size. The thick line represents the best-fit parameters obtained from our fitting (the spine streamer), while the thin lines represent the trajectories of the different bundle streamers. The blue and red colors represent the blueshifted and redshifted emission, respectively. \com{While the modeled trajectories of the bundle streamers fit reasonably well the observations, the PV diagrams show differences that Mendoza's model cannot account for.}

Finally, the ratio between the rotational energy (gas motions around an axis perpendicular to the equatorial disk plane, $v_{\phi}$) and the accretion energy (inward radial motions of the streamers, $v_r$) is estimated as $(v_{\phi}/v_r)^2$. For the four streamers, the average values of this ratio between 2,000~au and 3,500~au from the protostar \com{(an intermediate distance from the protostar, which avoids the rotation motions within the molecular \mfl{disk/envelope} and does not extend too much beyond the length of the St1 and St2 streamers, allowing a fair comparison between all of them)}, are 1.2, 0.3, 0.9, and 0.9, respectively. Except for St2, dominated by \com{more direct} inward motions, the rest show a balance between both components at these distances. 

\begin{table}[t!]
\centering
\caption{Parameters of the best fits of the four different streamers}
\begin{tabular}{c c c c c c c c}
\hline
\hline
Streamer & $\mu$ & $\nu$ & $\theta_0$ & $\Delta \theta_0$ & $\phi_0$ & $\Delta \phi_0$ & $e$ \\
 &  &  & $[^{\circ}]$ & $[^\circ]$& $[^{\circ}]$ & $[^\circ]$ & \\
\hline \\
St1 & 0.59 & 0.78 & 101.3 & $\pm10$ & 305.2 & $\pm10$ & 0.88 \\
St2 & 0.66 & 0.46 & 30.5 & $\pm15$ & 285.3 & $\pm15$  & 0.86 \\
St3 & 0.31 & $2\times10^{-3}$ & 116.7 & $\pm5$ & 194.2 & $\pm5$ & 0.75 \\
St4 & 0.55 & $2\times10^{-2}$ & 130.1 & $\pm25$ & 132.6 & $\pm25$ & 0.68 \\
\hline 
\end{tabular}
\label{tab:mendozaparameters}
\end{table}

\subsection{Mass and momentum within the streamers}
\label{subsec:masses}

In this section we estimate the column density and mass using the 1.2~mm continuum emission from the Band~6 image (lower panel of Figure \ref{fig:Bfields}), which shows that the dust emission also traces the streamers.  We also derive accretion rates using the lifetime for the streamers derived in the previous section \ref{subsec:streamers}. Let us make the caveat that the values derived for the column density are dependent on the adopted temperature and dust opacity. They may also be affected by missing flux.

We assume the dust emission is optically thin, and the dust grain  opacity $\kappa_{1.3mm}$ is 0.51~cm$^2$~g$^{-1}$ for dust with thin ice mantles \citep{1994Ossenkopf}, a spectral index $\beta=1.6$ for average grains of the Interstellar Medium \citep{2006Draine} \mfl{and also consistent with the average $\beta$ value found in a dust streamer associated to a Class I young star in the Orion Molecular Cloud \citep{2024Cacciapuoti}}, and a dust temperature ranging between 25~K and 70~K, as we found for the dust surrounding the MM2 outflow, the second main protostar in the GGD27 region \citep{2025LopezVazquez}. We also make the typical assumption that the gas-to-dut ratio is 100.  
 
Using the well known expression \citep[e.g., ][]{1983Hildebrand}: $$ M=\frac{R~d^2~S_{\nu}}{\kappa_{\nu}~B_{\nu}(T_{dust})}\quad,$$
where $S_{\nu}$, $d$, $\kappa_{\nu}$ and $B_{\nu}(T_{dust})$ are the total observed flux density, the distance, the grain opacity and the brightness of a blackbody at temperature $T_{dust}$.  We derive that the mass of the streamers range between 0.2\msun and 2.8\msun, and their column densities between $1.2\times10^{22}$ and $9.8\times10^{22}$~cm$^{-2}$ (Table \ref{tab:masses}). Using the lifetimes of the streamers, their approximate accretion rates (streamer mass~$/$~lifetime) are or order $10^{-4}$--$10^{-5}$\msun~yr$^{-1}$. Adding the 4 streamers together the total accretion rate would be of order $3\times10^{-4}$\msun~yr$^{-1}$ \citep[a rate consistent with those estimated for high-mass protostars,][]{2016Beltran}, indicating that in about $10^4$~yr $\sim3$\msun might be incorporated to the MM1 system.

\com{In an independent way we also derived values for the gas mass using the H$_2$CO ($4_{1,3}-3_{1,2}$) line emission. An inspection of the velocity cube reveals that the H$_2$CO ($4_{1,3}-3_{1,2}$) emission ranges from optically thin to moderate optically thick ($\tau\simeq1$) with an average opacity\footnote{Estimated using the expression $\tau_{\nu}=-log\{1-T_R/[J(T_{ex})-J(T_{bg}]\}$.} not greater than 0.4 within the streamers for excitation temperatures between 25~K and 70~K. We first computed column densities using the simplified equation 85 in \cite{2015Mangum}:
\begin{equation*}
    N_{H_2CO}=\frac{3h~Q_{rot}~exp\left(\frac{E_u}{kT_{ex}}\right)}{8~\pi^3~S\mu^2~g_u~\left( exp\left(\frac{h\nu}{kT_{ex}}\right) -1\right)}\int{\frac{T_R~dv}{J(T_{ex})-J(T_{bg})}}\quad ,
\end{equation*}
where $h$ and $k$ are the Planck and Boltzmann constants, $Q_{rot}$ is the partition function, $E_u$ and $g_u$ are the upper level energy and degeneracy, $T_{ex}$ and $T_{bg}$ are the excitation and background temperatures, $\nu$ is the transition rest frequency, $T_R$ is the radiation temperature, $d\nu$ is the line-width, and the Rayleigh-Jeans equivalent temperature is defined as $J(T)=h\nu/\left[k~\left(exp\left(h\nu/kT\right) - 1 \right)\right]$. Second, as in \cite{2025LopezVazquez}, we adopt a formaldehyde abundance X[H$_2$CO]$\approx10^{-10}$ \citep{Gerner2014,Tang2018,Gieser2021,Zhao2024}. Comparing the dust and the H$_2 $CO emission within each individual streamer (Figure \ref{fig:Bfields}) we see that the abundance can vary within a factor of 4. With all of this, and adopting a fiducial $T_{ex}=40~K$  and a streamer depth equivalent to its width,  the resulting masses are 0.2\msun, 0.7\msun, 0.2\msun and 3.0\msun, for the St1, St2, St3 and St4 streamers, respectively. Despite the uncertainties in temperature, depth, abundance and possible missing flux of the molecular interferometric maps, these values are in quite good agreement with the masses derived from the dust emission. 
}

\begin{table*}[t!]
\centering
\caption{Physical and dynamical characteristics of the streamers}
\begin{tabular}{l c c c c c c }
\hline
\hline
Streamer & $\Delta v$ & Mass & $N_{col}$ & n & Lifetime & $\dot{M}$\\
 & [\kms]  & [\msun] & [$10^{22}$~cm$^{-2}$]  & [$10^{6}$~cm$^{-3}$] & [yr] & [$10^{-5}$~\msun~yr$^{-1}$]  \\
 \hline \\
St1 & 3.5 & 0.1--0.5 & 1.2--4.0 & 0.4--1.5 & 5300 & 3 \\
St2 & 4.0 & 0.4--1.2 & 1.9--6.1 & 0.7--2.2 & 5500 & 1.3 \\ 
St3 & 4.0 & 0.2--0.7 & 2.4--7.8 & 0.9--2.8 & 29000 & 0.06 \\ 
St4 & 8.5 & 0.8--2.8 & 3.0--9.8 & 1.1--3.6 & 12300 & 25 \\ 
\hline
\end{tabular}
\tablefoot{The second column shows the velocity \com{range covering} each streamer, derived from the H$_2$CO emission. The rest of the columns show values derived from the analysis of the 1.2~mm continuum emission. The two values of each table entry are extracted by using dust temperatures of 70~K and 25~K, respectively. Lifetimes are derived numerically using the equations for the streamer models (Section \ref{subsec:streamers}).}
\label{tab:masses}
\end{table*}

\subsection{Magnetic strength along the streamers}
\label{subsec:bfield}

\begin{figure*}[ht!]
\centering
\subfigure{\includegraphics[width=0.75\linewidth]{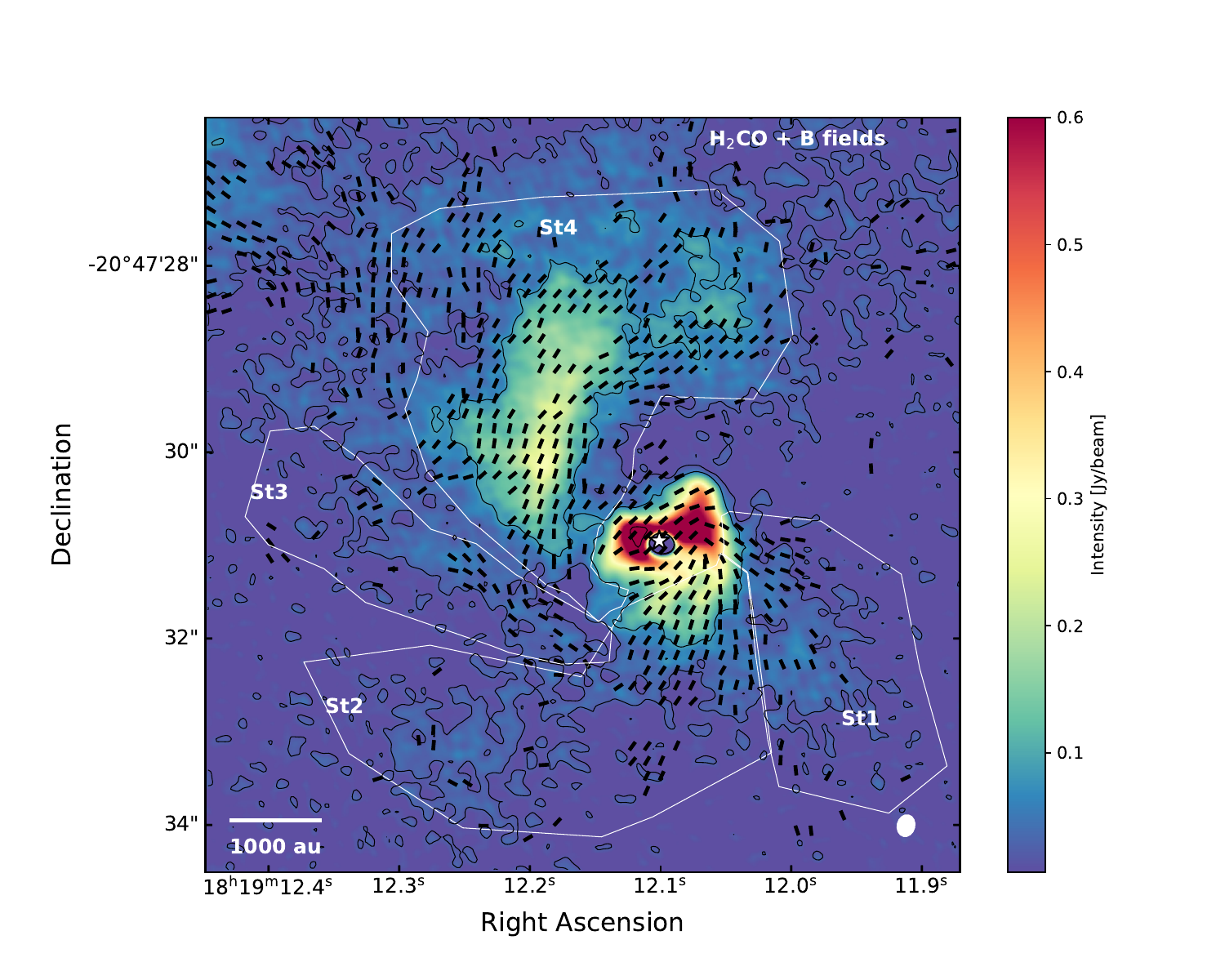}}
\subfigure{\includegraphics[width=0.75\linewidth]{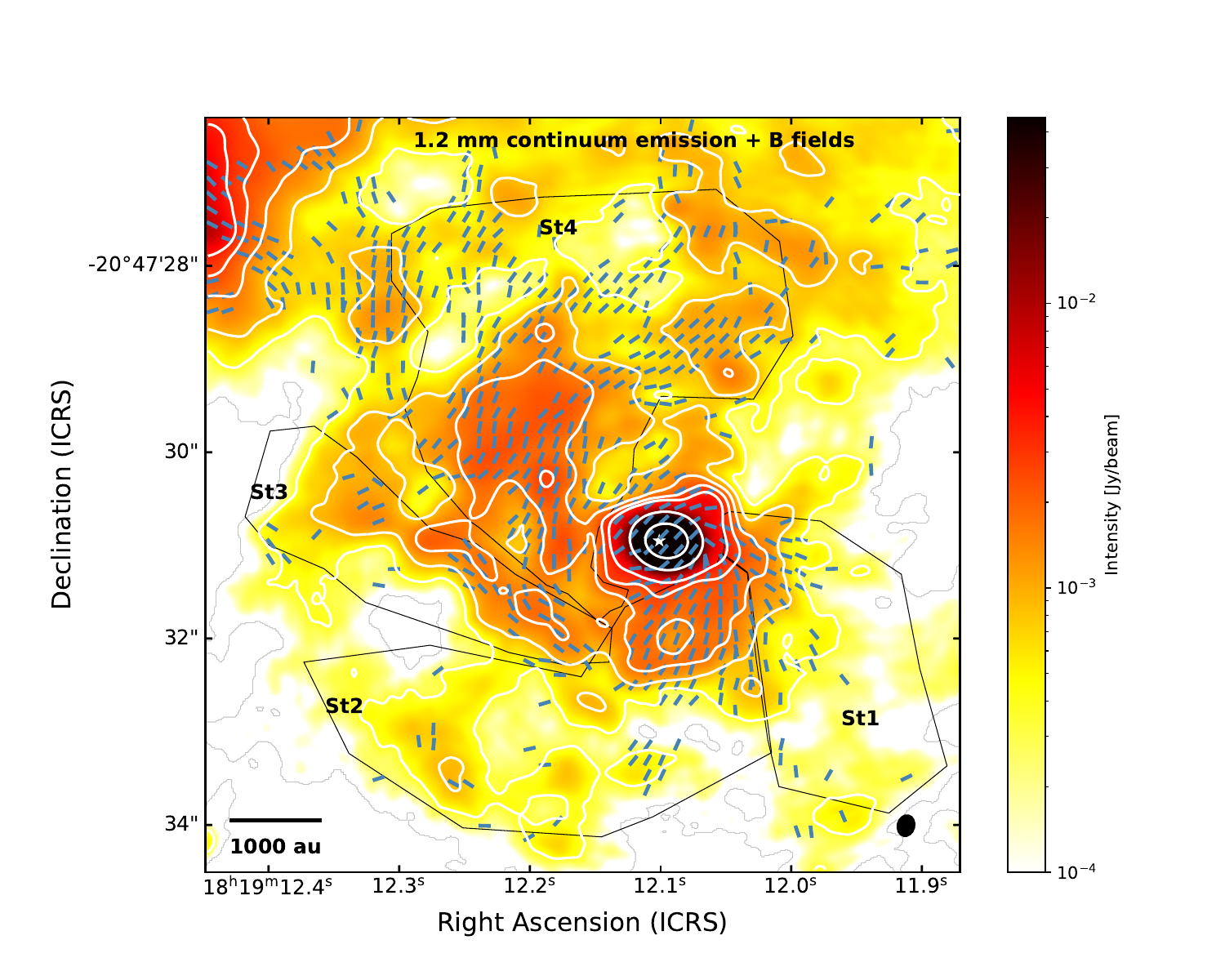}}
\caption{\textbf{Top:} H$_2$CO integrated intensity moment 0 image (color scale and contours) overlaid with polarization black half-vectors rotated by 90$\degr$, and hence displaying the magnetic field orientations, toward the GGD~27 region, focused on MM1 (position marked with a white star). Four regions --defined using the velocity cube emission-- with possible large-scale accretion streamers are marked with black polygons. The contour levels are displayed at 2, 8, 80, and 200 times $\sigma$, the rms noise level of $11$\mjy measured in a nearby region devoid of bright emission. The synthesized beam is in the bottom right corner. \textbf{Bottom:} 1.2~mm continuum emission (color scale and contours) overlaid with polarization blue half-vectors rotated by 90$\degr$. The contour levels start at 3, 7, 13, 21, 31, 190, and 1190 times $\sigma$, the rms noise level of $0.12$\mjy~\kms in the image. Symbols as in the upper panel.}
\label{fig:Bfields}
\end{figure*}

Figure \ref{fig:Bfields} shows the inferred direction of the magnetic field toward the GGD~27 star-forming complex. In the \com{streamers} identified in Section \ref{subsec:streamers}, the field orientations reasonably agree with the fitted trajectories of the gas \com{(within $30\degr$ for the most part of the streamer spines)}. We \com{derived} the strength of the magnetic fields in the plane of the sky associated with these structures using the DCF method. Since the overall morphology of the magnetic field toward the GGD~27 complex is curved, in accordance with the trajectory of the infalling gas toward MM1, we adopt an analogous approach as in \cite{2024Cortes}. First, we use the expression:
\begin{equation}\label{eq:bpos}
\mathrm{B_{\mathrm{pos}}\simeq\xi\frac{\sigma_{\mathrm{los}}}{\langle\sigma_{\phi}\rangle}\sqrt{4\pi\rho}\quad.}
\end{equation}
In this equation, $\mathrm{B_{\mathrm{pos}}}$ is the plane-of-the-sky component of the magnetic field strength, $\xi=1/2$ is a correction factor derived from simulations of turbulent clouds \citep{2001Ostriker}, $\mathrm{\sigma_{\mathrm{los}}}$ is the velocity dispersion along the line of sight, $\langle\sigma_{\phi}\rangle$ is the dispersion of the magnetic field angles, and $\rho$ is the density. We make the caveat that there may be more appropriate $\xi$ correction factors for cylinders and other structure geometries \citep{2021Liu}, however, since the 3D structure of gas and magnetic field is uncertain, we just use the original $\xi=1/2$ factor.  We select four regions that encompass the gas streamers detected in the velocity cube, considering emission with intensity above a 3$\sigma$ threshold. We trim them toward the center to avoid the MM1 \mfl{disk/envelope system.} These regions present dust continuum emission, H$_2$CO emission and significant polarized emission (hence, magnetic field orientations; see Figure \ref{fig:Bfields}). 

In Section \ref{subsec:masses} we report the mass, average column density and particle density in these regions.  We also assume that their depth is similar to the width of the structures where they connect to the \mfl{envelope} in a direction perpendicular to the main gas trajectories (Section \ref{subsec:streamers}). Hence, we estimate average densities of $\rho\sim7\times10^{-18}$~g~cm$^{-3}$ ($1.5\times10^{-6}$~cm$^{-3}$; Table \ref{tab:bfields}). Let us remind here that, due to uncertainties in the dust temperature and opacity the density estimates derived from the continuum emission have at least 20\% uncertainty.

\begin{figure*}[t!]
\centering
\includegraphics[width=0.75\linewidth]{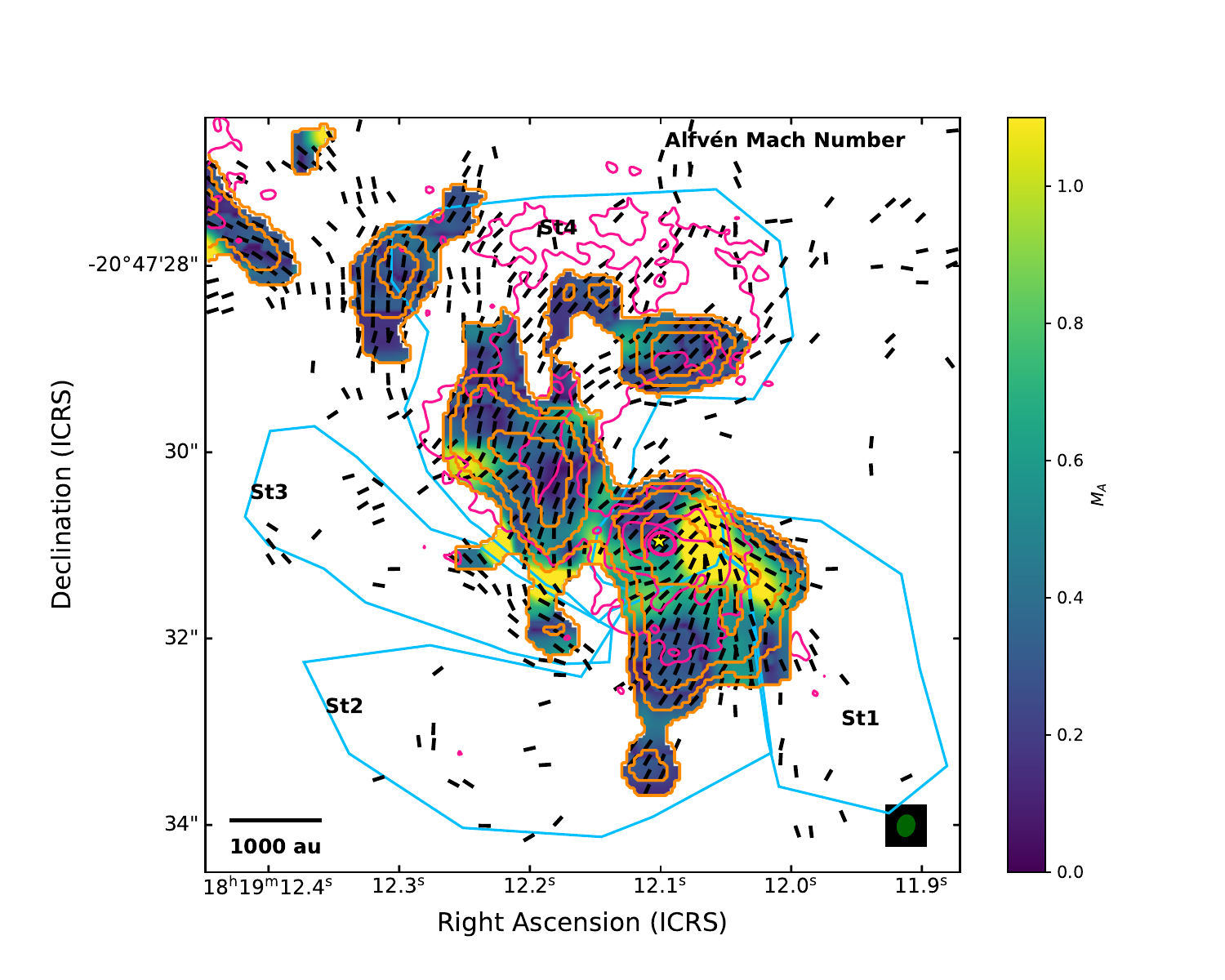}
\caption{Alfv\'enic Mach number, $\mathcal{M}_\mathcal{A}$, image. Orange contours mark different levels for the margin of error of the $\mathcal{M}_\mathcal{A}$ measurements to the 80\% confidence level (inner to outer contours mark margins of error of 0.19, 0.24, 0.28, and 0.33, respectively). Magenta contours show the integrated intensity image of the H$_2$CO emission (at 15, 50, 150, and 350 times the rms noise level of the peak emission). The synthesized beam of the molecular image is shown at the bottom right corner, along with a $16\times16$~pixel$^2$ black square, the size of the moving window used to compute the dispersions of the magnetic field orientations (black half-vectors). The four streamer regions are delineated by blue polygons. The yellow star marks the position of the GGD27--MM1 continuum peak.}
\label{fig:machA}
\end{figure*}

We extract the mean velocity dispersion from the  H$_2$CO emission, which is optically thin for most parts of the four structures \com{(for excitation temperatures between 25~K and 70~K, a detailed inspection of the velocity cube reveals that the H$_2$CO ($4_{1,3}-3_{1,2}$) emission ranges from optically thin to moderate optically thick, $\tau\simeq1$, with just a few small zones of optically thick emission)}. 
After channel width (0.5\kms) deconvolution, we subtract in quadrature the thermal component of the line, $\mathrm{\sigma_{\mathrm{los}}=\sqrt{\sigma^2_{\mathrm{observed}}-\sigma^2_{\mathrm{thermal}}}}$. This thermal component is derived using $\mathrm{\sigma_{\mathrm{thermal}}=\sqrt{(k_B T)/(\mu m_H)}}$, where $\mathrm{k_B}$ is the Boltzmann constant, $m_H$ is the atomic hydrogen weight, $T$ is the temperature of the gas and $\mu=30.031$ is the weight of the H$_2$CO molecule. 
At 40~K, $\mathrm{\sigma_{\mathrm{thermal}}=0.10}$\kms. The choice of temperature in the probed range between 25~K and 70~K has little impact --less than a few percent, well within the quoted uncertainties-- on the final magnetic field measurements, $\mathrm{B_{\mathrm{pos}}}$, and hence we adopt 40~K as an intermediate fiducial value.

\mfl{One possible complication in the usage of the DCF method can be contamination of the line profiles by large-scale gas motions, which may affect the derivation of the $\mathrm{\sigma_{\mathrm{los}}}$. However, within the size of the streamers, the individual line profiles (with independent measures every half beam, about 100~au) are not significantly widened due to overall velocity gradients. Indeed, considering a streamer length of $\sim6000$~au, for velocity gradients $\lesssim4$\kms, the possible line-widening contamination may be $\lesssim0.06$\kms ($\mathrm{\Delta V\cdot half\_beam/length}$), which is of the order of the statistical uncertainty in the $\mathrm{\sigma_{\mathrm{los}}}$ values (Table \ref{tab:bfields}). Also, taking into account that the gradients become steeper closer to the potential well, this rough estimate for the linewidth contamination can be thought as an upper limit in most parts of the streamers.}

As stated before, to derive the dispersion of the magnetic field angles, $\langle\sigma_{\phi}\rangle$, we use the same approach as in \cite{2024Cortes}, related to that used in \cite{2017Pattle} and \cite{2021Hwang}. We calculate the standard deviation within a moving window of size $0\farcs43$, which encompasses about 4 independent measurements per window \com{($2\times2$ beams)} and a margin of error of $\gtrsim4\degr$ with a confidence level of 80\%, reasonable for a quantitative analysis of the measured values of angle dispersion $\sigma_{\phi}$. 
After this, we calculate the mean of the standard deviations ($\langle\sigma_{\phi}\rangle$) within the moving windows for each of the four streamers. 
Then we plug these values in Equation \ref{eq:bpos} to derive the magnetic field strength in the plane of the sky. 
The uncertainty of the $\langle\sigma_{\phi}\rangle$ measurements is derived as the root mean square of the individual $\sigma_{\phi}$ measured for every moving window of a region, divided by the term $\mathrm{ \sqrt{2(N_{\mathrm{angle}}-1)~N_{\mathrm{window}}}}$, where $\mathrm{ N_{\mathrm{angle}}}$ is the number of independent angle measurements within one window and $\mathrm{N_{\mathrm{window}}}$ is the number of windows within one region. 
In Table \ref{tab:bfields}, we summarize the average values of $\mathrm{B_{\mathrm{pos}}}$ encountered for each of the selected regions. The mean plane-of-the-sky magnetic field strength in these regions is $\sim1.5$\,mG. Given the high uncertainty, dominated by the density estimate, and the relatively small statistics over the magnetic field half-vectors, $\mathrm{B_{\mathrm{pos}}}$ values (with uncertainties between 15\% and 25\%) need to be utilized with caution.
We also estimate the Alfv\'en speed, $v_{\mathcal A}$ and the Alfv\'en Mach number, \malfven. The Alfv\'en speed is $v_{\mathcal A}=B_{3D}/\sqrt{4\pi\rho}\approx(4/\pi)\cdot B_{\mathrm{pos}}/\sqrt{4\pi\rho}$, where we take $B_{3D}\approx(4/\pi)\cdot B_{\mathrm{pos}}$ \cite[see ][]{2004Crutcher}, despite this \com{$B_{3D}$} is a geometrical factor to derive statistical average values, and not so appropriate for individual $B$-fields with large inclinations with respect to the plane of the sky. \malfven can aid in diagnosing the relative balance between turbulent and magnetically dominated gas motions (see Section \ref{sec:discussion}). $\mathcal{M_A}=v_{\rm rms}/v_{\mathcal A}$ only depends on $\langle\sigma_{\phi}\rangle$ \com{\citep[and the inclination of the field][]{2024Cortes}}, since $v_{\rm rms}=\sqrt{3}\sigma_{\rm los}$ for isotropic turbulence, and the $\sigma_{\rm los}$ term \com{(the non-thermal velocity dispersion)} cancels with that implicit in the $B_{\mathrm{pos}}$ definition (Equation \ref{eq:bpos}), leaving $\mathcal{M_A}\approx[\sqrt{3}\pi/(4\xi)]\cdot\langle\sigma_{\phi}\rangle\approx\langle\sigma_{\phi}\rangle/21\fdg06$.
\com{Within the constraints of this} analysis, all the streamers are, on average, \com{sub-alfv\'enic}, which indicates a dominant magnetic field over the turbulence, \com{as expected for structures with a coherent bulk motion}. 

\begin{table*}[t!]
    \centering
    \caption{Magnetic field strengths in the plane of the sky, $B_{\mathrm{pos}}$, derived in the streamer-like regions, by using the DCF method.}
    \begin{tabular}{c c c c c c c c c c}
    \hline
    \hline
ID & Depth* & $n$\dag & $\rho$\dag & $\sigma_{\mathrm{los}}$ & $\mathrm{\mathcal{M}}$ & $<\sigma_{\phi}>$ & $B_{\mathrm{pos}}$ & $v_{\mathcal A}$ & $\mathrm{\mathcal{M_A}}$ \\
 & [au] & [$10^5$~cm$^{-3}$] & [$10^{-18}$~g.cm$^{-3}$] & [\kms] & & [$\degr$] & [mG] & [\kms] &  \\
\hline \\
St1 & 2380 & 8$\pm$2 & 3.7$\pm$0.7     & 0.26$\pm$0.03 & 1.2$\pm$0.3 & 8.5$\pm$0.9 & 0.45$\pm$0.08 & 0.8$\pm$0.2 & 0.54$\pm$0.04 \\
St2 & 2380 & 13$\pm$3 & 6$\pm$1     & 0.47$\pm$0.03 & 2.2$\pm$0.6 & 8.2$\pm$0.7 & 1.1$\pm$0.2 & 1.6$\pm$0.3 & 0.51$\pm$0.03 \\
St3 & 1260 & 22$\pm$4 & 10$\pm$2     & 0.55$\pm$0.05 & 2.6$\pm$0.6 & 11$\pm$1 & 1.3$\pm$0.2 & 1.5$\pm$0.3 & 0.64$\pm$0.05 \\
St4 & 3500 & 34$\pm$7 & 8$\pm$2    & 1.37$\pm$0.06 & 6$\pm$2 & 9.5$\pm$0.3 & 3.3$\pm$0.4 & 4.2$\pm$0.5 & 0.57$\pm$0.01 \\
\hline 
\end{tabular}
\tablefoot{* The depth is assumed to be the same as the transversal size of the considered structures.}
\tablefoot{\dag The densities are estimated under the assumption that the 3D-morphology of these regions is cylindrical. The uncertainties are estimated to be 30\% of the fiducial values used.}
\label{tab:bfields}
\end{table*}

\section{Discussion}
\label{sec:discussion}
In Section \ref{subsec:streamers} we fit the trajectories of the putative streamers feeding MM1's \mfl{disk/envelope}, while in Section \ref{subsec:bfield} we perform a statistical analysis to derive estimates for the average magnetic field strength of each streamer. In this Section, we evaluate the balance between different agents affecting gas motion within the streamers. For the St4 streamer, we suspect that the values derived may be affected by the overlapping of more than one structure in the line of sight. This is suggested by the larger line-widths found in this streamer (about 3 times larger than the other streamers, Table \ref{tab:masses}) \com{and the finding of double-peaked spectral profiles toward that region. The larger line-widths might} propagate into a larger turbulence velocity and magnetic field strength (see Table \ref{tab:bfields}). In what follows, we disregard the results for the St4 streamer.

First, we test if gas motions are thermally dominated. The speed of sound of the gas is $c_s=\sqrt{(k_B T)/(\mu m_H)}$, where we use $\mu=2.37$ as the mean molecular weight of the gas. For $T=40$~K this expression gives $c_s=0.37$\kms.  The majority of the molecular gas in the St2 and St3 streamers is supersonic (see weight-averaged Mach number, $\mathrm{\mathcal{M}}$, in Table \ref{tab:bfields}), while in the St1 streamer 68\% of the gas is supersonic and 22\% transonic. Hence, most of the gas in the streamers is supersonic and turbulence along the line-of-sight beam-size is non-thermally dominated.

Second, we evaluate if the magnetic field can overcome turbulence within the streamers. Figure \ref{fig:machA} shows the spatial distribution of the Alfv\'enic Mach number, \malfven, derived in places with good enough statistics of the polarization orientations. \com{The figure was built using the expression in section \ref{subsec:bfield} and deriving the dispersion of magnetic field orientations within a moving window of $16\times16$ pixel$^2$, in locations where the number of measurements complied with the criterion based on the confidence level of 80\% previously stated (at least 240 pixels worth of data).} Table \ref{tab:bfields} contains average values of \malfven for each streamer. Both the image and the statistical analysis show that the gas is preferentially sub-alfv\'enic, indicating that the magnetic forces dominate the turbulent motions on average. However, there are locations with large \malfven values (dark zones in Figure \ref{fig:machA}). Most of them lie within the disk \mfl{and envelope} of the MM1 protostellar system and part of the St1 streamer. 
Other \malfven$\gtrsim1$ regions are due to abrupt changes in polarization orientations, which break the DCF assumption \com{of straight line orientation of the magnetic field within one of the analyzed} moving window. 

Now, to assess the role of magnetic fields against other forces we will compare the relative magnitudes of the terms in the virial equation \citep{1953Chandrasekhar,2004Stahler,2021Sanhueza} for each streamer. \com{Note that instead of energy terms we use specific energies (i.e., energy per unit mass).} In particular, we estimate:
\begin{itemize}
    \item The thermal plus turbulent energy: $e_{\rm turb}=3\sigma_{\rm los}^2$, where we use the averaged observed velocity dispersion. Note that the turbulence contemplated by this term is measured within a beam.
    \item The bulk kinetic energy (rotation, infall, expansion and large-scale turbulence): $e_K=1/2~v^2_{\rm bulk}$. We estimate the bulk velocity by setting an upper limit of 2.25\kms. This value is retrieved from the maximum shift in the line-of-sight velocity with respect to the systemic velocity in the pv-diagrams of the streamers between 2,000\,au and 3500\,au (i.e., between $1\farcs5$--$2\farcs5$; see Figure \ref{fig:pvdiagrams}). Within this term, there could be contributions from rotation, infall (caused by gravity), but also by external pressure produced by the HH~80/81 jet or any other source of external large-scale turbulence.
    \item The magnetic energy: $e_B=B^2/(8\pi\rho)$. We make use of the average magnetic field values derived in Section \ref{subsec:bfield}.
    \item The gravitational energy \com{(we use its absolute value here)}: $e_G=GM_{\rm central}/R$. We assume a total mass (protostar plus disk plus envelope) of 30\msun, and again, we probe regions at distances 2,000--3,500\,au from the MM1 protostar position.
\end{itemize}
Let us make the caveat that we are considering mean spatial values for all these quantities, despite they vary with the distance to the protostar. Here we just give \com{an estimate of} their comparison. 
Taking all of this into account we find for the St1, St2 and St3 streamers: a ratio of magnetic to turbulent energy $e_B/e_{turb}\approx1.2-1.8$ (in line with the gas being sub-alfv\'enic), magnetic to kinetic energy $e_B/e_{K}\approx0.15-0.40$, magnetic to gravitational $e_B/e_G\approx0.03-0.1$, and kinetic to gravitational $e_K/e_G\approx0.25$. Plus, a rough estimate of the virial parameter gives $\alpha_{vir,B,rot}=-(2~e_K+e_B)/e_G\approx0.5-0.6$. If these estimates are qualitatively correct, at intermediate scales in the streamers (at about 3,000\,au from the star) gravity dominates the energy budget: the kinetic energy is $\sim e_G/4$ (almost equally distributed in rotation and accretion, see Section \ref{subsec:streamers}), the magnetic energy is $\sim e_G/10$, and the turbulent energy is $\sim e_G/20$. Hence, the magnetic fields oppose relatively little to the gravity pull, whereas the kinetic energy, which may have a contribution from the expansion of the HH\,80/81 jet, has a more relevant role.

The scope of this paper does not include determining the actual shape of the streamers. Despite naively thought as cylinder-like structures due to  their filamentary appearance, current MHD simulations have produced gravomagneto sheetlets during the collapse of molecular cloud cores \citep{2024Tu}. These sheetlets may be regarded as blankets of denser material feeding gas and carrying angular momentum to the \mfl{disk/envelope}, while orbiting the potential well.
In our estimation of the mass and density (Section \ref{subsec:masses}) we have assumed the depth of the streamers to be the same as their width close to the envelope, despite not knowing anything about it. If we take instead, half of the width, the density will be twice and the magnetic field may be a factor of $\sqrt{2}$ larger, provided that the geometric constant factor $\xi$ does not change as well. On the other hand, $v_{\mathcal A}$, $\mathrm{\mathcal{M_A}}$, and the specific energies do not depend on the density and our qualitative analysis stays the same.
A putative sheetlet appearance for the streamers is much closer to the look presented by the ALMA H$_2$CO observations \citep[compare Figure \ref{fig:Bfields} and Figures 2(h) and 2(k) in][]{2024Tu}). After fitting the trajectory of a single particle to the selected data for each streamer, we plot the trajectories of particles whose initial positions describe a circular grid surrounding the original one. This builds a cylinder-like pipe of particles collapsing to the \mfl{disk/envelope system}. While this bundle of trajectories may not be the best 3D-representation of the streamers, it results in a more complete characterization in the position-position-velocity space. However, even considering the whole bundle trajectories both the morphology and kinematics cannot be perfectly matched by Mendoza's model (Figure \ref{fig:pvdiagrams}), which suggests that other agents such as magnetic fields and external pressure (e.g., outflowing gas) have a non-negligible contribution to the gas motions. We estimated in Section \ref{subsec:streamers} the rotation and infall energies are comparable. Hence, depending on the level of external pressure, magnetic energy could be up to $\sim0.5$ times the rotational energy, which may affect the trajectories from Mendoza's model by removing angular momentum \citep[as recently reported by][]{2026Huang}.

Throughout this work, and also in the literature in general, the streamers linked to disks and envelopes are regarded as accretion structures. Simulations show that some of them can eject at least part of their material if portions of their structure have enough angular momentum. As a first approximation, let us assume this is not the case in GGD\,27 and all the gas in the streamers is falling toward MM1. However, some of the streamers seem to hover in projection the position of other protostars in the protocluster. In particular, the northern St4 streamer spreads over the position of protostars ALMA~4, ALMA~9, and ALMA~11, in projection \citep{Busquet2019}. Also, the furthest end to MM1 of the St3 streamer is not far from MM2 (about $\sim2\arcsec$ apart from that position, about 2800~au). If the gas from the streamers was connected to these protostars, it seems that the larger gravitational pull from MM1 has overcome the local gravitational wells from the rest of the cluster members and is dragging material from their surroundings.

\section{Summary of results}
\label{sec:conclusions}
We conducted 1.0~mm and 1.2~mm ALMA observations (including molecular and polarized continuum emission) toward the massive star-forming region GGD~27. We identified four large-scale (up to 7,000\,au in length) accretion streamers apparently collapsing on to the envelope/disk of the GGD~27--MM1 protostellar system. We fitted their trajectories using the prescriptions of the so called Mendoza's model, and calculated their masses (typically a fraction of a solar mass), densities and accretion rates (of order $10^{-4}$\msun~yr$^{-1}$ in total). We used a statistical approach based on the classical DCF analysis to derive average magnetic field strengths for each streamer. This resulted in strengths around 1.5~mG. Our Virial analysis lead to conclude that the streamers are gravitationally bound at distances $\sim3,000$\,au from the central protostar. The magnetic energy in the streamers is $\sim1/10$th of the gravitational energy, $\sim2/5$th of the kinetic energy, and $\sim 3/2$ times the turbulent energy. While the gas is supersonic and sub-alfv\'enic, the rotational and infalling motions induced by gravity seem to dominate the dynamics of the streamers. There could be another kinetic contribution (expansion from the HH\,80/81 \com{jet} or the action of magnetic field forces) which make Mendoza's model trajectories not a perfect match for the line of sight velocities observed. In addition, we found some indications of a possible competitive accretion in the region, with MM1 dominating the gravitational well over the other nearby members of the cluster.

\begin{acknowledgments}
    M.F.L., J.M.G and G.B. acknowledge support from the PID2023-146675NB-I00 (MCI-AEI-FEDER, UE) program. This work was also partly supported by the Spanish program Unidad de Excelencia María de Maeztu CEX2020-001058-M, financed by MCIN/AEI/10.13039/501100011033, and by the MaX-CSIC Excellence Award MaX4-SOMMA-IC.
    J.A.L.-V. and C.-F. Lee acknowledge a grant  from the National Science and Technology Council of Taiwan (NSTC 112–2112–M–001–039–MY3). 
    L.A.Z. acknowledges financial support from CONACyT-280775, UNAM-PAPIIT IN110618, and IN112323 grants, México.
    M.T.B. acknowledges financial support through the INAF Large Grant {\it The role of MAGnetic fields in MAssive star formation} (MAGMA). 
    P.S. was partially supported by a Grant-in-Aid for Scientific Research (KAKENHI No JP24K17100) of the Japan Society for the Promotion of Science (JSPS).\\
    This paper makes use of the following ALMA data: ADS/JAO.ALMA\# 2015.1.00480.S. ALMA is a partnership of ESO (representing its member states), NSF (USA) and NINS (Japan), together with NRC (Canada), MOST and ASIAA (Taiwan), and KASI (Republic of Korea), in cooperation with the Republic of Chile. The Joint ALMA Observatory is operated by ESO, AUI/NRAO and NAOJ.
\end{acknowledgments}

\bibliographystyle{aa} 
\bibliography{sample631} 

@ARTICLE{2024Gupta,
       author = {{Gupta}, Aashish and {Miotello}, Anna and {Williams}, Jonathan P. and {Birnstiel}, Til and {Kuffmeier}, Michael and {Yen}, Hsi-Wei},
        title = "{TIPSY: Trajectory of Infalling Particles in Streamers around Young stars. Dynamical analysis of the streamers around S CrA and HL Tau}",
      journal = {\aap},
         year = 2024,
        month = mar,
       volume = {683},
          eid = {A133},
        pages = {A133},
          doi = {10.1051/0004-6361/202348007},
archivePrefix = {arXiv},
       eprint = {2401.10403},
 primaryClass = {astro-ph.SR},
       adsurl = {https://ui.adsabs.harvard.edu/abs/2024A&A...683A.133G}
}

@ARTICLE{2020Pineda,
       author = {{Pineda}, Jaime E. and {Segura-Cox}, Dominique and {Caselli}, Paola and {Cunningham}, Nichol and {Zhao}, Bo and {Schmiedeke}, Anika and {Maureira}, Mar{\'\i}a Jos{\'e} and {Neri}, Roberto},
        title = "{A protostellar system fed by a streamer of 10,500 au length}",
      journal = {Nat. Astron.},
         year = 2020,
        month = jan,
       volume = {4},
        pages = {1158-1163},
          doi = {10.1038/s41550-020-1150-z},
archivePrefix = {arXiv},
       eprint = {2007.13430},
 primaryClass = {astro-ph.GA},
       adsurl = {https://ui.adsabs.harvard.edu/abs/2020NatAs...4.1158P}
}

@ARTICLE{2022ValdiviaMena,
       author = {{Valdivia-Mena}, M.~T. and {Pineda}, J.~E. and {Segura-Cox}, D.~M. and {Caselli}, P. and {Neri}, R. and {L{\'o}pez-Sepulcre}, A. and {Cunningham}, N. and {Bouscasse}, L. and {Semenov}, D. and {Henning}, Th. and {Pi{\'e}tu}, V. and {Chapillon}, E. and {Dutrey}, A. and {Fuente}, A. and {Guilloteau}, S. and {Hsieh}, T.~H. and {Jim{\'e}nez-Serra}, I. and {Marino}, S. and {Maureira}, M.~J. and {Smirnov-Pinchukov}, G.~V. and {Tafalla}, M. and {Zhao}, B.},
        title = "{PRODIGE - envelope to disk with NOEMA. I. A 3000 au streamer feeding a Class I protostar}",
      journal = {\aap},
         year = 2022,
        month = nov,
       volume = {667},
          eid = {A12},
        pages = {A12},
          doi = {10.1051/0004-6361/202243310},
archivePrefix = {arXiv},
       eprint = {2208.01023},
 primaryClass = {astro-ph.GA},
       adsurl = {https://ui.adsabs.harvard.edu/abs/2022A&A...667A..12V}
}

@ARTICLE{2024Cacciapuoti,
       author = {{Cacciapuoti}, L. and {Macias}, E. and {Gupta}, A. and {Testi}, L. and {Miotello}, A. and {Espaillat}, C. and {K{\"u}ffmeier}, M. and {van Terwisga}, S. and {Tobin}, J. and {Grant}, S. and {Manara}, C.~F. and {Segura-Cox}, D. and {Wendeborn}, J. and {Klessen}, R.~S. and {Maury}, A.~J. and {Lebreuilly}, U. and {Hennebelle}, P. and {Molinari}, S.},
        title = "{A dusty streamer infalling onto the disk of a class I protostar. ALMA dual-band constraints on grain properties and the mass-infall rate}",
      journal = {\aap},
         year = 2024,
        month = feb,
       volume = {682},
          eid = {A61},
        pages = {A61},
          doi = {10.1051/0004-6361/202347486},
archivePrefix = {arXiv},
       eprint = {2311.13723},
 primaryClass = {astro-ph.EP},
       adsurl = {https://ui.adsabs.harvard.edu/abs/2024A&A...682A..61C}
}

@ARTICLE{2023Kuffmeier,
       author = {{Kuffmeier}, Michael and {Jensen}, Sigurd S. and {Haugb{\o}lle}, Troels},
        title = "{Rejuvenating infall: a crucial yet overlooked source of mass and angular momentum}",
      journal = {Eur. Phys. J. Plus},
         year = 2023,
        month = mar,
       volume = {138},
       number = {3},
          eid = {272},
        pages = {272},
          doi = {10.1140/epjp/s13360-023-03880-y},
archivePrefix = {arXiv},
       eprint = {2303.05261},
 primaryClass = {astro-ph.SR},
       adsurl = {https://ui.adsabs.harvard.edu/abs/2023EPJP..138..272K}
}

@ARTICLE{1983Hildebrand,
       author = {{Hildebrand}, R.~H.},
        title = "{The determination of cloud masses and dust characteristics from submillimetre thermal emission.}",
      journal = {\qjras},
         year = 1983,
        month = sep,
       volume = {24},
        pages = {267-282},
       adsurl = {https://ui.adsabs.harvard.edu/abs/1983QJRAS..24..267H}
}

@article{2026Hwang,
doi = {10.3847/1538-3881/ae18c9},
url = {https://doi.org/10.3847/1538-3881/ae18c9},
year = {2025},
month = {dec},
publisher = {The American Astronomical Society},
volume = {171},
number = {1},
pages = {50},
author = {Hwang, Jihye and Sanhueza, Patricio and Girart, Josep Miquel and Stephens, Ian W. and Beltrán, Maria T. and Law, Chi Yan and Zhang, Qizhou and Liu, Junhao and Cortés, Paulo and Olguin, Fernando A. and Koch, Patrick M. and Nakamura, Fumitaka and Saha, Piyali and Wang, Jia-Wei and Xu, Fengwei and Beuther, Henrik and Morii, Kaho and Fernández López, Manuel and Jiao, Wenyu and Kim, Kee-Tae and Li, Shanghuo and Zapata, Luis A. and Kim, Jongsoo and Choudhury, Spandan and Cheng, Yu and Pattle, Kate and Eswaraiah, Chakali and Sandhyarani, Panigrahy and Dewangan, L. K. and Jadhav, O. R.},
title = {Magnetic Fields in Massive Star-forming Regions (MagMaR). VI. Magnetic Field Dragging in the Filamentary High-mass Star-forming Region G35.20–0.74N Due to Gravity},
journal = {\apj}
}

@ARTICLE{2018Vig,
       author = {{Vig}, S. and {Veena}, V.~S. and {Mandal}, S. and {Tej}, A. and {Ghosh}, S.~K.},
        title = "{Detection of non-thermal emission from the massive protostellar jet HH80-81 at low radio frequencies using GMRT}",
      journal = {\mnras},
         year = 2018,
        month = mar,
       volume = {474},
       number = {3},
        pages = {3808-3816},
          doi = {10.1093/mnras/stx3032},
archivePrefix = {arXiv},
       eprint = {1711.07642},
 primaryClass = {astro-ph.SR},
       adsurl = {https://ui.adsabs.harvard.edu/abs/2018MNRAS.474.3808V}
}

@ARTICLE{2026Fedriani,
       author = {{Fedriani}, R. and {Anglada}, G. and {o Garatti}, A. Caratti and {G{\'o}mez}, J.~F. and {Masqu{\'e}}, J. and {Osorio}, M. and {Stecklum}, B. and {Rodr{\'\i}guez-Kamenetzky}, A.~R. and {Galv{\'a}n-Madrid}, R. and {Carrasco-Gonz{\'a}lez}, C. and {Bl{\'a}zquez-Calero}, G. and {Placinta-Mitrea}, A.~F. and {Sanna}, A. and {Cesaroni}, R. and {Moscadelli}, L. and {Ray}, T.~P. and {Coffey}, D. and {Fuller}, G.~A.},
        title = "{Diverse stages of star formation in the IRAS 18162-2048 region: Emergence of UV feedback}",
      journal = {\aap},
         year = 2026,
        month = mar,
       volume = {708},
          eid = {A11},
        pages = {A11},
          doi = {10.1051/0004-6361/202558460},
archivePrefix = {arXiv},
       eprint = {2512.07604},
 primaryClass = {astro-ph.GA},
       adsurl = {https://ui.adsabs.harvard.edu/abs/2026A&A...708A..11F}
}

@ARTICLE{2026Gupta,
       author = {{Gupta}, Aashish and {Hales}, Antonio S. and {Cleeves}, L. Ilsedore and {Alves}, Felipe and {Bhowmik}, Trisha and {Cuello}, Nicol{\'a}s and {Girart}, Josep M. and {Li}, Zhi-Yun and {Miotello}, Anna and {Zhu}, Zhaohuan and {Zurlo}, Alice},
        title = "{A Tale of Three Tails: A Misaligned Streamer and Mysterious Structures around [BHB2007]-1}",
      journal = {\apj},
         year = 2026,
        month = feb,
       volume = {998},
       number = {1},
          eid = {105},
        pages = {105},
          doi = {10.3847/1538-4357/ae2477},
archivePrefix = {arXiv},
       eprint = {2512.00295},
 primaryClass = {astro-ph.SR},
       adsurl = {https://ui.adsabs.harvard.edu/abs/2026ApJ...998..105G}
}

@ARTICLE{2025Cortes,
       author = {{Cort{\'e}s}, P.~C. and {Pineda}, J.~E. and {Hsieh}, T.-H. and {Tobin}, J.~J. and {Saha}, P. and {Girart}, J.~M. and {Le Gouellec}, V.~J.~M. and {Stephens}, I.~W. and {Looney}, L.~W. and {Koumpia}, E. and {Valdivia-Mena}, M.~T. and {Cacciapuoti}, L. and {Gieser}, C. and {Offner}, S.~S.~R. and {Caselli}, P. and {Sanhueza}, P. and {Segura-Cox}, D. and {Fern{\'a}ndez-L{\'o}pez}, M. and {Morii}, K. and {Huang}, B. and {Alves}, F.~O. and {Zhang}, Q. and {Kwon}, W. and {Hull}, C.~L.~H. and {Li}, Z.-Y.},
        title = "{First Results from ALPPS: A Sub-Alfv{\'e}nic Streamer in SVS 13A}",
      journal = {\apjl},
         year = 2025,
        month = oct,
       volume = {992},
       number = {2},
          eid = {L31},
        pages = {L31},
          doi = {10.3847/2041-8213/ae0c04},
archivePrefix = {arXiv},
       eprint = {2509.21701},
 primaryClass = {astro-ph.GA},
       adsurl = {https://ui.adsabs.harvard.edu/abs/2025ApJ...992L..31C}
}

@article{2026Huang,
  author  = {Huang, Bo and Girart, Josep M. and Stephens, Ian W. and Megeath, Tom and Le Gouellec, Valentin J. M. and Murillo, Nadia M. and Cortés, Paulo and Fernández-López, Manuel and Li, Zhi-Yun and Myers, Philip C. and others},
  title   = {Magnetogravitationally regulated streamer accretion onto a class 0 protostellar system},
  journal = {Sci. Adv.},
  volume  = {12},
  number  = {21},
  year    = {2026},
  month   = {may},
  doi     = {10.1126/sciadv.aec0413},
  note    = {Open access}
}

@ARTICLE{2017Pattle,
       author = {{Pattle}, Kate and {Ward-Thompson}, Derek and {Berry}, David and {Hatchell}, Jennifer and {Chen}, Huei-Ru and {Pon}, Andy and {Koch}, Patrick M. and {Kwon}, Woojin and {Kim}, Jongsoo and {Bastien}, Pierre and {Cho}, Jungyeon and {Coud{\'e}}, Simon and {Di Francesco}, James and {Fuller}, Gary and {Furuya}, Ray S. and {Graves}, Sarah F. and {Johnstone}, Doug and {Kirk}, Jason and {Kwon}, Jungmi and {Lee}, Chang Won and {Matthews}, Brenda C. and {Mottram}, Joseph C. and {Parsons}, Harriet and {Sadavoy}, Sarah and {Shinnaga}, Hiroko and {Soam}, Archana and {Hasegawa}, Tetsuo and {Lai}, Shih-Ping and {Qiu}, Keping and {Friberg}, Per},
        title = "{The JCMT BISTRO Survey: The Magnetic Field Strength in the Orion A Filament}",
      journal = {\apj},
         year = 2017,
        month = sep,
       volume = {846},
       number = {2},
          eid = {122},
        pages = {122},
          doi = {10.3847/1538-4357/aa80e5},
archivePrefix = {arXiv},
       eprint = {1707.05269},
 primaryClass = {astro-ph.GA},
       adsurl = {https://ui.adsabs.harvard.edu/abs/2017ApJ...846..122P}
}

@ARTICLE{2021Hwang,
       author = {{Hwang}, Jihye and {Kim}, Jongsoo and {Pattle}, Kate and {Kwon}, Woojin and {Sadavoy}, Sarah and {Koch}, Patrick M. and {Hull}, Charles L.~H. and {Johnstone}, Doug and {Furuya}, Ray S. and {Won Lee}, Chang and {Arzoumanian}, Doris and {Tahani}, Mehrnoosh and {Eswaraiah}, Chakali and {Liu}, Tie and {Kirchschlager}, Florian and {Kim}, Kee-Tae and {Tamura}, Motohide and {Kwon}, Jungmi and {Lyo}, A.-Ran and {Soam}, Archana and {Kang}, Ji-hyun and {Bourke}, Tyler L. and {Matsumura}, Masafumi and {Mairs}, Steve and {Kim}, Gwanjeong and {Park}, Geumsook and {Nakamura}, Fumitaka and {Onaka}, Takashi and {Tang}, Xindi and {Liu}, Hong-Li and {Ward-Thompson}, Derek and {Li}, Di and {Hoang}, Thiem and {Hasegawa}, Tetsuo and {Qiu}, Keping and {Lai}, Shih-Ping and {Bastien}, Pierre},
        title = "{The JCMT BISTRO Survey: The Distribution of Magnetic Field Strengths toward the OMC-1 Region}",
      journal = {\apj},
         year = 2021,
        month = jun,
       volume = {913},
       number = {2},
          eid = {85},
        pages = {85},
          doi = {10.3847/1538-4357/abf3c4},
archivePrefix = {arXiv},
       eprint = {2103.16144},
 primaryClass = {astro-ph.GA},
       adsurl = {https://ui.adsabs.harvard.edu/abs/2021ApJ...913...85H}
}

@ARTICLE{2025Olguin,
       author = {{Olguin}, Fernando A. and {Sanhueza}, Patricio and {Ginsburg}, Adam and {Chen}, Huei-Ru Vivien and {Tanaka}, Kei E.~I. and {Lu}, Xing and {Morii}, Kaho and {Nakamura}, Fumitaka and {Li}, Shanghuo and {Cheng}, Yu and {Zhang}, Qizhou and {Luo}, Qiuyi and {Oya}, Yoko and {Sakai}, Takeshi and {Saito}, Masao and {Guzm{\'a}n}, Andr{\'e}s E.},
        title = "{Massive extended streamers feed high-mass young stars}",
      journal = {Sci. Adv.},
         year = 2025,
        month = aug,
       volume = {11},
       number = {34},
          eid = {eadw4512},
        pages = {eadw4512},
          doi = {10.1126/sciadv.adw4512},
archivePrefix = {arXiv},
       eprint = {2508.15889},
 primaryClass = {astro-ph.GA},
       adsurl = {https://ui.adsabs.harvard.edu/abs/2025SciA...11.4512O}
}

@ARTICLE{2016Beltran,
       author = {{Beltr{\'a}n}, M.~T. and {de Wit}, W.~J.},
        title = "{Accretion disks in luminous young stellar objects}",
      journal = {\aapr},
         year = 2016,
        month = jan,
       volume = {24},
          eid = {6},
        pages = {6},
          doi = {10.1007/s00159-015-0089-z},
archivePrefix = {arXiv},
       eprint = {1509.08335},
 primaryClass = {astro-ph.GA},
       adsurl = {https://ui.adsabs.harvard.edu/abs/2016A&ARv..24....6B}
}

@ARTICLE{2025Beuther,
       author = {{Beuther}, H. and {Olguin}, F.~A. and {Sanhueza}, P. and {Cunningham}, N. and {Ginsburg}, A.},
        title = "{Hierarchical accretion flow from the G351 infrared dark filament to its central cores}",
      journal = {\aap},
         year = 2025,
        month = mar,
       volume = {695},
          eid = {A51},
        pages = {A51},
          doi = {10.1051/0004-6361/202452754},
archivePrefix = {arXiv},
       eprint = {2502.13866},
 primaryClass = {astro-ph.GA},
       adsurl = {https://ui.adsabs.harvard.edu/abs/2025A&A...695A..51B}
}

@ARTICLE{2025Liu,
       author = {{Liu}, X.-C. and {van Dishoeck}, E.~F. and {Hogerheijde}, M.~R. and {van Gelder}, M.~L. and {Chen}, Y. and {Liu}, T. and {van't Hoff}, M. and {Drozdovskaya}, M.~N. and {Artur de la Villarmois}, E. and {Mai}, X.-F. and {Tychoniec}, {\L}.},
        title = "{Sulfur oxides tracing streamers and shocks at low-mass protostellar disk{\textendash}envelope interfaces}",
      journal = {\aap},
         year = 2025,
        month = sep,
       volume = {701},
          eid = {A141},
        pages = {A141},
          doi = {10.1051/0004-6361/202554186},
archivePrefix = {arXiv},
       eprint = {2507.22870},
 primaryClass = {astro-ph.GA},
       adsurl = {https://ui.adsabs.harvard.edu/abs/2025A&A...701A.141L}
}

@ARTICLE{2022Lu,
       author = {{Lu}, Xing and {Li}, Guang-Xing and {Zhang}, Qizhou and {Lin}, Yuxin},
        title = "{A massive Keplerian protostellar disk with flyby-induced spirals in the Central Molecular Zone}",
      journal = {Nat. Astron.},
         year = 2022,
        month = may,
       volume = {6},
        pages = {837-843},
          doi = {10.1038/s41550-022-01681-4},
archivePrefix = {arXiv},
       eprint = {2206.00202},
 primaryClass = {astro-ph.GA},
       adsurl = {https://ui.adsabs.harvard.edu/abs/2022NatAs...6..837L}
}

@ARTICLE{2004Crutcher,
       author = {{Crutcher}, Richard M. and {Nutter}, D.~J. and {Ward-Thompson}, D. and {Kirk}, J.~M.},
        title = "{SCUBA Polarization Measurements of the Magnetic Field Strengths in the L183, L1544, and L43 Prestellar Cores}",
      journal = {\apj},
         year = 2004,
        month = jan,
       volume = {600},
       number = {1},
        pages = {279-285},
          doi = {10.1086/379705},
archivePrefix = {arXiv},
       eprint = {astro-ph/0305604},
 primaryClass = {astro-ph},
       adsurl = {https://ui.adsabs.harvard.edu/abs/2004ApJ...600..279C}
}

@ARTICLE{2006Vaillancourt,
       author = {{Vaillancourt}, John E.},
        title = "{Placing Confidence Limits on Polarization Measurements}",
      journal = {\pasp},
         year = 2006,
        month = sep,
       volume = {118},
       number = {847},
        pages = {1340-1343},
          doi = {10.1086/507472},
archivePrefix = {arXiv},
       eprint = {astro-ph/0603110},
 primaryClass = {astro-ph},
       adsurl = {https://ui.adsabs.harvard.edu/abs/2006PASP..118.1340V}
}

@ARTICLE{2021Olguin,
       author = {{Olguin}, Fernando A. and {Sanhueza}, Patricio and {Guzm{\'a}n}, Andr{\'e}s E. and {Lu}, Xing and {Saigo}, Kazuya and {Zhang}, Qizhou and {Silva}, Andrea and {Chen}, Huei-Ru Vivien and {Li}, Shanghuo and {Ohashi}, Satoshi and {Nakamura}, Fumitaka and {Sakai}, Takeshi and {Wu}, Benjamin},
        title = "{Digging into the Interior of Hot Cores with ALMA (DIHCA). I. Dissecting the High-mass Star-forming Core G335.579-0.292 MM1}",
      journal = {\apj},
         year = 2021,
        month = mar,
       volume = {909},
       number = {2},
          eid = {199},
        pages = {199},
          doi = {10.3847/1538-4357/abde3f},
archivePrefix = {arXiv},
       eprint = {2101.08284},
 primaryClass = {astro-ph.GA},
       adsurl = {https://ui.adsabs.harvard.edu/abs/2021ApJ...909..199O}
}

@ARTICLE{2024Zapata,
       author = {{Zapata}, Luis A. and {Fern{\'a}ndez-L{\'o}pez}, Manuel and {Sanhueza}, Patricio and {Girart}, Josep M. and {Rodr{\'\i}guez}, Luis F. and {Cort{\'e}s}, Paulo and {Koch}, Patrick and {Beltr{\'a}n}, Maria T. and {Pattle}, Kate and {Beuther}, Henrik and {Saha}, Piyali and {Jiao}, Wenyu and {Xu}, Fengwei and {Lu}, Xing Walker and {Olguin}, Fernando and {Li}, Shanghuo and {Stephens}, Ian W. and {Kang}, Ji-hyun and {Cheng}, Yu and {Choudhury}, Spandan and {Morii}, Kaho and {Chung}, Eun Jung and {Wang}, Jia-Wei and {Hwang}, Jihye and {Lyo}, A. -Ran and {Zhang}, Q. and {Chen}, Huei-Ru Vivien},
        title = "{Magnetic Fields in Massive Star-forming Regions (MagMaR). IV. Tracing the Magnetic Fields in the O-type Protostellar System IRAS 16547{\textendash}4247}",
      journal = {\apj},
         year = 2024,
        month = oct,
       volume = {974},
       number = {2},
          eid = {257},
        pages = {257},
          doi = {10.3847/1538-4357/ad701d},
archivePrefix = {arXiv},
       eprint = {2408.10199},
 primaryClass = {astro-ph.SR},
       adsurl = {https://ui.adsabs.harvard.edu/abs/2024ApJ...974..257Z}
}

@ARTICLE{2024Saha,
       author = {{Saha}, Piyali and {Sanhueza}, Patricio and {Padovani}, Marco and {Girart}, Josep M. and {Cort{\'e}s}, Paulo C. and {Morii}, Kaho and {Liu}, Junhao and {S{\'a}nchez-Monge}, {\'A}. and {Galli}, Daniele and {Basu}, Shantanu and {Koch}, Patrick M. and {Beltr{\'a}n}, Maria T. and {Li}, Shanghuo and {Beuther}, Henrik and {Stephens}, Ian W. and {Nakamura}, Fumitaka and {Zhang}, Qizhou and {Jiao}, Wenyu and {Fern{\'a}ndez-L{\'o}pez}, M. and {Hwang}, Jihye and {Chung}, Eun Jung and {Pattle}, Kate and {Zapata}, Luis A. and {Xu}, Fengwei and {Olguin}, Fernando A. and {Kang}, Ji-hyun and {Karoly}, Janik and {Law}, Chi-Yan and {Wang}, Jia-Wei and {Csengeri}, Timea and {Lu}, Xing and {Cheng}, Yu and {Kim}, Jongsoo and {Choudhury}, Spandan and {Chen}, Huei-Ru Vivien and {Hull}, Charles L.~H.},
        title = "{Magnetic Fields in Massive Star-forming Regions (MagMaR): Unveiling an Hourglass Magnetic Field in G333.46{\textendash}0.16 Using ALMA}",
      journal = {\apjl},
         year = 2024,
        month = sep,
       volume = {972},
       number = {1},
          eid = {L6},
        pages = {L6},
          doi = {10.3847/2041-8213/ad660c},
archivePrefix = {arXiv},
       eprint = {2407.16654},
 primaryClass = {astro-ph.GA},
       adsurl = {https://ui.adsabs.harvard.edu/abs/2024ApJ...972L...6S}
}

@ARTICLE{2024Cortes,
       author = {{Cort{\'e}s}, Paulo C. and {Girart}, Josep M. and {Sanhueza}, Patricio and {Liu}, Junhao and {Mart{\'\i}n}, Sergio and {Stephens}, Ian W. and {Beuther}, Henrik and {Koch}, Patrick M. and {Fern{\'a}ndez-L{\'o}pez}, M. and {S{\'a}nchez-Monge}, {\'A}lvaro and {Wang}, Jia-Wei and {Morii}, Kaho and {Li}, Shanghuo and {Saha}, Piyali and {Zhang}, Qizhou and {Rebolledo}, David and {Zapata}, Luis A. and {Kang}, Ji-hyun and {Jiao}, Wenyu and {Kim}, Jongsoo and {Cheng}, Yu and {Hwang}, Jihye and {Chung}, Eun Jung and {Choudhury}, Spandan and {Lyo}, A. -Ran and {Olguin}, Fernando},
        title = "{MagMaR III{\textemdash}Resisting the Pressure, Is the Magnetic Field Overwhelmed in NGC6334I?}",
      journal = {\apj},
         year = 2024,
        month = sep,
       volume = {972},
       number = {1},
          eid = {115},
        pages = {115},
          doi = {10.3847/1538-4357/ad59a7},
archivePrefix = {arXiv},
       eprint = {2406.14663},
 primaryClass = {astro-ph.GA},
       adsurl = {https://ui.adsabs.harvard.edu/abs/2024ApJ...972..115C}
}

@ARTICLE{2021Cortes,
       author = {{Cort{\'e}s}, Paulo C. and {Sanhueza}, Patricio and {Houde}, Martin and {Mart{\'\i}n}, Sergio and {Hull}, Charles L.~H. and {Girart}, Josep M. and {Zhang}, Qizhou and {Fernandez-Lopez}, Manuel and {Zapata}, Luis A. and {Stephens}, Ian W. and {Li}, Hua-bai and {Wu}, Benjamin and {Olguin}, Fernando and {Lu}, Xing and {Guzm{\'a}n}, Andres E. and {Nakamura}, Fumitaka},
        title = "{Magnetic Fields in Massive Star-forming Regions (MagMaR). II. Tomography through Dust and Molecular Line Polarization in NGC 6334I(N)}",
      journal = {\apj},
         year = 2021,
        month = dec,
       volume = {923},
       number = {2},
          eid = {204},
        pages = {204},
          doi = {10.3847/1538-4357/ac28a1},
archivePrefix = {arXiv},
       eprint = {2109.09270},
 primaryClass = {astro-ph.GA},
       adsurl = {https://ui.adsabs.harvard.edu/abs/2021ApJ...923..204C}
}

@ARTICLE{2025Sanhueza,
       author = {{Sanhueza}, Patricio and {Liu}, Junhao and {Morii}, Kaho and {Girart}, Josep Miquel and {Zhang}, Qizhou and {Stephens}, Ian W. and {Jackson}, James M. and {Cort{\'e}s}, Paulo C. and {Koch}, Patrick M. and {Cyganowski}, Claudia J. and {Saha}, Piyali and {Beuther}, Henrik and {Zhang}, Suinan and {Beltr{\'a}n}, Maria T. and {Cheng}, Yu and {Olguin}, Fernando A. and {Lu}, Xing and {Choudhury}, Spandan and {Pattle}, Kate and {Fern{\'a}ndez-L{\'o}pez}, Manuel and {Hwang}, Jihye and {Kang}, Ji-hyun and {Karoly}, Janik and {Ginsburg}, Adam and {Lyo}, A. -Ran and {Taniguchi}, Kotomi and {Jiao}, Wenyu and {Eswaraiah}, Chakali and {Luo}, Qiu-yi and {Wang}, Jia-Wei and {Commer{\c{c}}on}, Beno{\^\i}t and {Li}, Shanghuo and {Xu}, Fengwei and {Chen}, Huei-Ru Vivien and {Zapata}, Luis A. and {Chung}, Eun Jung and {Nakamura}, Fumitaka and {Panigrahy}, Sandhyarani and {Sakai}, Takeshi},
        title = "{Magnetic Fields in Massive Star-forming Regions (MagMaR). V. The Magnetic Field at the Onset of High-mass Star Formation}",
      journal = {\apj},
         year = 2025,
        month = feb,
       volume = {980},
       number = {1},
          eid = {87},
        pages = {87},
          doi = {10.3847/1538-4357/ad9d40},
archivePrefix = {arXiv},
       eprint = {2412.08790},
 primaryClass = {astro-ph.GA},
       adsurl = {https://ui.adsabs.harvard.edu/abs/2025ApJ...980...87S}
}

@ARTICLE{1951Davis,
       author = {{Davis}, Leverett},
        title = "{The Strength of Interstellar Magnetic Fields}",
      journal = {Phys. Rev.},
         year = 1951,
        month = mar,
       volume = {81},
       number = {5},
        pages = {890-891},
          doi = {10.1103/PhysRev.81.890.2},
       adsurl = {https://ui.adsabs.harvard.edu/abs/1951PhRv...81..890D}
}

@ARTICLE{2023Olguin,
       author = {{Olguin}, Fernando A. and {Sanhueza}, Patricio and {Chen}, Huei-Ru Vivien and {Lu}, Xing and {Oya}, Yoko and {Zhang}, Qizhou and {Ginsburg}, Adam and {Taniguchi}, Kotomi and {Li}, Shanghuo and {Morii}, Kaho and {Sakai}, Takeshi and {Nakamura}, Fumitaka},
        title = "{Digging into the Interior of Hot Cores with ALMA: Spiral Accretion into the High-mass Protostellar Core G336.01-0.82}",
      journal = {\apjl},
         year = 2023,
        month = dec,
       volume = {959},
       number = {2},
          eid = {L31},
        pages = {L31},
          doi = {10.3847/2041-8213/ad1100},
archivePrefix = {arXiv},
       eprint = {2311.18006},
 primaryClass = {astro-ph.GA},
       adsurl = {https://ui.adsabs.harvard.edu/abs/2023ApJ...959L..31O}
}

@ARTICLE{2023Xu,
       author = {{Xu}, Feng-Wei and {Wang}, Ke and {Liu}, Tie and {Goldsmith}, Paul F. and {Zhang}, Qizhou and {Juvela}, Mika and {Liu}, Hong-Li and {Qin}, Sheng-Li and {Li}, Guang-Xing and {Tej}, Anandmayee and {Garay}, Guido and {Bronfman}, Leonardo and {Li}, Shanghuo and {Wu}, Yue-Fang and {G{\'o}mez}, Gilberto C. and {V{\'a}zquez-Semadeni}, Enrique and {Tatematsu}, Ken'ichi and {Ren}, Zhiyuan and {Zhang}, Yong and {Toth}, L. Viktor and {Liu}, Xunchuan and {Yue}, Nannan and {Zhang}, Siju and {Baug}, Tapas and {Issac}, Namitha and {Stutz}, Amelia M. and {Liu}, Meizhu and {Fuller}, Gary A. and {Tang}, Mengyao and {Zhang}, Chao and {Dewangan}, Lokesh and {Lee}, Chang Won and {Zhou}, Jianwen and {Xie}, Jinjin and {Jiao}, Wenyu and {Wang}, Chao and {Liu}, Rong and {Luo}, Qiuyi and {Soam}, Archana and {Eswaraiah}, Chakali},
        title = "{ATOMS: ALMA Three-millimeter Observations of Massive Star-forming regions - XV. Steady accretion from global collapse to core feeding in massive hub-filament system SDC335}",
      journal = {\mnras},
         year = 2023,
        month = apr,
       volume = {520},
       number = {3},
        pages = {3259-3285},
          doi = {10.1093/mnras/stad012},
archivePrefix = {arXiv},
       eprint = {2301.01895},
 primaryClass = {astro-ph.GA},
       adsurl = {https://ui.adsabs.harvard.edu/abs/2023MNRAS.520.3259X}
}

@ARTICLE{2010Keto,
       author = {{Keto}, Eric and {Zhang}, Qizhou},
        title = "{The standard model of star formation applied to massive stars: accretion discs and envelopes in molecular lines}",
      journal = {\mnras},
         year = 2010,
        month = jul,
       volume = {406},
       number = {1},
        pages = {102-111},
          doi = {10.1111/j.1365-2966.2010.16672.x},
archivePrefix = {arXiv},
       eprint = {1002.1864},
 primaryClass = {astro-ph.GA},
       adsurl = {https://ui.adsabs.harvard.edu/abs/2010MNRAS.406..102K}
}

@ARTICLE{2006Keto,
       author = {{Keto}, Eric and {Wood}, Kenneth},
        title = "{Observations on the Formation of Massive Stars by Accretion}",
      journal = {\apj},
         year = 2006,
        month = feb,
       volume = {637},
       number = {2},
        pages = {850-859},
          doi = {10.1086/498611},
archivePrefix = {arXiv},
       eprint = {astro-ph/0510176},
 primaryClass = {astro-ph},
       adsurl = {https://ui.adsabs.harvard.edu/abs/2006ApJ...637..850K}
}

@ARTICLE{2021Liu,
       author = {{Liu}, Junhao and {Zhang}, Qizhou and {Commer{\c{c}}on}, Beno{\^\i}t and {Valdivia}, Valeska and {Maury}, Ana{\"e}lle and {Qiu}, Keping},
        title = "{Calibrating the Davis-Chandrasekhar-Fermi Method with Numerical Simulations: Uncertainties in Estimating the Magnetic Field Strength from Statistics of Field Orientations}",
      journal = {\apj},
         year = 2021,
        month = oct,
       volume = {919},
       number = {2},
          eid = {79},
        pages = {79},
          doi = {10.3847/1538-4357/ac0cec},
archivePrefix = {arXiv},
       eprint = {2106.09934},
 primaryClass = {astro-ph.GA},
       adsurl = {https://ui.adsabs.harvard.edu/abs/2021ApJ...919...79L}
}

@ARTICLE{2024Tu,
       author = {{Tu}, Yisheng and {Li}, Zhi-Yun and {Zhu}, Zhaohuan and {Hsu}, Chun-Yen},
        title = "{Fragmentation of dense rotation-dominated structures fed by collapsing gravo-magneto-sheetlets and origin of misaligned 100 au-scale binaries and multiple systems}",
      journal = {\mnras},
         year = 2024,
        month = aug,
       volume = {532},
       number = {3},
        pages = {3135-3150},
          doi = {10.1093/mnras/stae1639},
archivePrefix = {arXiv},
       eprint = {2403.07777},
 primaryClass = {astro-ph.SR},
       adsurl = {https://ui.adsabs.harvard.edu/abs/2024MNRAS.532.3135T}
}

@BOOK{2004Stahler,
       author = {{Stahler}, Steven W. and {Palla}, Francesco},
        title = "{The Formation of Stars}",
        year = 2004,
        publisher = {Wiley-VCH},
    	address = {Weinheim, Germany},
    	isbn = {978-3527405596},
       adsurl = {https://ui.adsabs.harvard.edu/abs/2004fost.book.....S}
}

@ARTICLE{2001Ostriker,
       author = {{Ostriker}, Eve C. and {Stone}, James M. and {Gammie}, Charles F.},
        title = "{Density, Velocity, and Magnetic Field Structure in Turbulent Molecular Cloud Models}",
      journal = {\apj},
         year = 2001,
        month = jan,
       volume = {546},
       number = {2},
        pages = {980-1005},
          doi = {10.1086/318290},
archivePrefix = {arXiv},
       eprint = {astro-ph/0008454},
 primaryClass = {astro-ph},
       adsurl = {https://ui.adsabs.harvard.edu/abs/2001ApJ...546..980O}
}

@ARTICLE{1953Chandrasekhar,
       author = {{Chandrasekhar}, S. and {Fermi}, E.},
        title = "{Magnetic Fields in Spiral Arms.}",
      journal = {\apj},
         year = 1953,
        month = jul,
       volume = {118},
        pages = {113},
          doi = {10.1086/145731},
       adsurl = {https://ui.adsabs.harvard.edu/abs/1953ApJ...118..113C}
}

@ARTICLE{2023FernandezLopez,
       author = {{Fern{\'a}ndez-L{\'o}pez}, M. and {Girart}, J.~M. and {L{\'o}pez-V{\'a}zquez}, J.~A. and {Estalella}, R. and {Busquet}, G. and {Curiel}, S. and {A{\~n}ez-L{\'o}pez}, N.},
        title = "{Disk and Envelope Streamers of the GGD 27-MM1 Massive Protostar}",
      journal = {\apj},
         year = 2023,
        month = oct,
       volume = {956},
       number = {2},
          eid = {82},
        pages = {82},
          doi = {10.3847/1538-4357/ace786},
archivePrefix = {arXiv},
       eprint = {2307.06178},
 primaryClass = {astro-ph.GA},
       adsurl = {https://ui.adsabs.harvard.edu/abs/2023ApJ...956...82F}
}

@ARTICLE{2021Sanhueza,
       author = {{Sanhueza}, Patricio and {Girart}, Josep Miquel and {Padovani}, Marco and {Galli}, Daniele and {Hull}, Charles L.~H. and {Zhang}, Qizhou and {Cortes}, Paulo and {Stephens}, Ian W. and {Fern{\'a}ndez-L{\'o}pez}, Manuel and {Jackson}, James M. and {Frau}, Pau and {Kock}, Patrick M. and {Wu}, Benjamin and {Zapata}, Luis A. and {Olguin}, Fernando and {Lu}, Xing and {Silva}, Andrea and {Tang}, Ya-Wen and {Sakai}, Takeshi and {Guzm{\'a}n}, Andr{\'e}s E. and {Tatematsu}, Ken'ichi and {Nakamura}, Fumitaka and {Chen}, Huei-Ru Vivien},
        title = "{Gravity-driven Magnetic Field at  1000 au Scales in High-mass Star Formation}",
      journal = {\apjl},
         year = 2021,
        month = jul,
       volume = {915},
       number = {1},
          eid = {L10},
        pages = {L10},
          doi = {10.3847/2041-8213/ac081c},
archivePrefix = {arXiv},
       eprint = {2106.03866},
 primaryClass = {astro-ph.GA},
       adsurl = {https://ui.adsabs.harvard.edu/abs/2021ApJ...915L..10S}
}

@ARTICLE{2021FernandezLopez,
       author = {{Fern{\'a}ndez-L{\'o}pez}, M. and {Sanhueza}, P. and {Zapata}, L.~A. and {Stephens}, I. and {Hull}, C. and {Zhang}, Q. and {Girart}, J.~M. and {Koch}, P.~M. and {Cort{\'e}s}, P. and {Silva}, A. and {Tatematsu}, K. and {Nakamura}, F. and {Guzm{\'a}n}, A.~E. and {Nguyen Luong}, Q. and {Guzm{\'a}n Ccolque}, E. and {Tang}, Y. -W. and {Chen}, H. -R.~V.},
        title = "{Magnetic Fields in Massive Star-forming Regions (MagMaR). I. Linear Polarized Imaging of the Ultracompact H II Region G5.89-0.39}",
      journal = {\apj},
         year = 2021,
        month = may,
       volume = {913},
       number = {1},
          eid = {29},
        pages = {29},
          doi = {10.3847/1538-4357/abf2b6},
archivePrefix = {arXiv},
       eprint = {2104.03331},
 primaryClass = {astro-ph.GA},
       adsurl = {https://ui.adsabs.harvard.edu/abs/2021ApJ...913...29F}
}

@ARTICLE{2025LopezVazquez,
       author = {{L{\'o}pez-V{\'a}zquez}, J.~A. and {Fern{\'a}ndez-L{\'o}pez}, M. and {Girart}, J.~M. and {Curiel}, S. and {Estalella}, R. and {Busquet}, G. and {Zapata}, L.~A. and {Lee}, C. -F. and {Galv{\'a}n-Madrid}, R.},
        title = "{Erosion of a dense molecular core by a strong outflow from a massive protostar}",
      journal = {\aap},
         year = 2025,
        month = mar,
       volume = {695},
          eid = {A236},
        pages = {A236},
          doi = {10.1051/0004-6361/202453196},
archivePrefix = {arXiv},
       eprint = {2502.17786},
 primaryClass = {astro-ph.SR},
       adsurl = {https://ui.adsabs.harvard.edu/abs/2025A&A...695A.236L}
}

@ARTICLE{2025RodriguezKamenetzky,
       author = {{Rodr{\'\i}guez-Kamenetzky}, A. and {Pasetto}, A. and {Carrasco-Gonz{\'a}lez}, C. and {Rodr{\'\i}guez}, L.~F. and {G{\'o}mez}, J.~L. and {Anglada}, G. and {Torrelles}, J.~M. and {Gomes}, N.~R.~C. and {Vig}, S. and {Mart{\'\i}}, J.},
        title = "{Helical Magnetic Field in a Massive Protostellar Jet}",
      journal = {\apjl},
         year = 2025,
        month = jan,
       volume = {978},
       number = {2},
          eid = {L31},
        pages = {L31},
          doi = {10.3847/2041-8213/ad9b26},
archivePrefix = {arXiv},
       eprint = {2501.07622},
 primaryClass = {astro-ph.GA},
       adsurl = {https://ui.adsabs.harvard.edu/abs/2025ApJ...978L..31R}
}

@ARTICLE{1991Aspin,
       author = {{Aspin}, C. and {McCaughrean}, M.~J. and {Casali}, M.~M. and {Geballe}, T.~R.},
        title = "{Near-IR imaging and spectroscopy of GGD 27-IRS.}",
      journal = {\aap},
         year = 1991,
        month = dec,
       volume = {252},
        pages = {299},
       adsurl = {https://ui.adsabs.harvard.edu/abs/1991A&A...252..299A}
}

@ARTICLE{2001Bonnell,
       author = {{Bonnell}, I.~A. and {Bate}, M.~R. and {Clarke}, C.~J. and {Pringle}, J.~E.},
        title = "{Competitive accretion in embedded stellar clusters}",
      journal = {\mnras},
         year = 2001,
        month = may,
       volume = {323},
       number = {4},
        pages = {785-794},
          doi = {10.1046/j.1365-8711.2001.04270.x},
archivePrefix = {arXiv},
       eprint = {astro-ph/0102074},
 primaryClass = {astro-ph},
       adsurl = {https://ui.adsabs.harvard.edu/abs/2001MNRAS.323..785B}
}

@INPROCEEDINGS{2022Pineda,
       author = {{Pineda}, J.~E. and {Arzoumanian}, D. and {Andre}, P. and {Friesen}, R.~K. and {Zavagno}, A. and {Clarke}, S.~D. and {Inoue}, T. and {Chen}, C. and {Lee}, Y. and {Soler}, J.~D. and {Kuffmeier}, M.},
        title = "{From Bubbles and Filaments to Cores and Disks: Gas Gathering and Growth of Structure Leading to the Formation of Stellar Systems}",
    booktitle = {Protostars and Planets VII},
         year = 2023,
       editor = {{Inutsuka}, S. and {Aikawa}, Y. and {Muto}, T. and {Tomida}, K. and {Tamura}, M.},
       series = {Astronomical Society of the Pacific Conference Series},
       volume = {534},
        month = jul,
        pages = {233},
          doi = {10.48550/arXiv.2205.03935},
archivePrefix = {arXiv},
       eprint = {2205.03935},
 primaryClass = {astro-ph.GA},
       adsurl = {https://ui.adsabs.harvard.edu/abs/2023ASPC..534..233P}
}

@ARTICLE{2010Carrasco,
       author = {{Carrasco-Gonz{\'a}lez}, Carlos and {Rodr{\'\i}guez}, Luis F. and {Anglada}, Guillem and {Mart{\'\i}}, Josep and {Torrelles}, Jos{\'e} M. and {Osorio}, Mayra},
        title = "{A Magnetized Jet from a Massive Protostar}",
      journal = {Sci},
         year = 2010,
        month = nov,
       volume = {330},
       number = {6008},
        pages = {1209},
          doi = {10.1126/science.1195589},
archivePrefix = {arXiv},
       eprint = {1011.6254},
 primaryClass = {astro-ph.GA},
       adsurl = {https://ui.adsabs.harvard.edu/abs/2010Sci...330.1209C}
}

@ARTICLE{2012Carrasco,
       author = {{Carrasco-Gonz{\'a}lez}, Carlos and {Galv{\'a}n-Madrid}, Roberto and {Anglada}, Guillem and {Osorio}, Mayra and {D'Alessio}, Paola and {Hofner}, Peter and {Rodr{\'\i}guez}, Luis F. and {Linz}, Hendrik and {Araya}, Esteban D.},
        title = "{Resolving the Circumstellar Disk around the Massive Protostar Driving the HH 80-81 Jet}",
      journal = {\apjl},
         year = 2012,
        month = jun,
       volume = {752},
       number = {2},
          eid = {L29},
        pages = {L29},
          doi = {10.1088/2041-8205/752/2/L29},
archivePrefix = {arXiv},
       eprint = {1205.3302},
 primaryClass = {astro-ph.GA},
       adsurl = {https://ui.adsabs.harvard.edu/abs/2012ApJ...752L..29C}
}

@ARTICLE{2006Draine,
       author = {{Draine}, B.~T.},
        title = "{On the Submillimeter Opacity of Protoplanetary Disks}",
      journal = {\apj},
         year = 2006,
        month = jan,
       volume = {636},
       number = {2},
        pages = {1114-1120},
          doi = {10.1086/498130},
archivePrefix = {arXiv},
       eprint = {astro-ph/0507292},
 primaryClass = {astro-ph},
       adsurl = {https://ui.adsabs.harvard.edu/abs/2006ApJ...636.1114D}
}

@ARTICLE{2011FernandezLopez2,
       author = {{Fern{\'a}ndez-L{\'o}pez}, M. and {Girart}, J.~M. and {Curiel}, S. and {G{\'o}mez}, Y. and {Ho}, P.~T.~P. and {Patel}, N.},
        title = "{A Rotating Molecular Disk Toward IRAS 18162-2048, the Exciting Source of HH 80-81}",
      journal = {\aj},
         year = 2011,
        month = oct,
       volume = {142},
       number = {4},
          eid = {97},
        pages = {97},
          doi = {10.1088/0004-6256/142/4/97},
archivePrefix = {arXiv},
       eprint = {1107.3176},
 primaryClass = {astro-ph.GA},
       adsurl = {https://ui.adsabs.harvard.edu/abs/2011AJ....142...97F}
}

@ARTICLE{1994Ossenkopf,
       author = {{Ossenkopf}, V. and {Henning}, Th.},
        title = "{Dust opacities for protostellar cores.}",
      journal = {\aap},
         year = 1994,
        month = nov,
       volume = {291},
        pages = {943-959},
       adsurl = {https://ui.adsabs.harvard.edu/abs/1994A&A...291..943O}
}

@ARTICLE{2011FernandezLopez1,
       author = {{Fern{\'a}ndez-L{\'o}pez}, M. and {Curiel}, S. and {Girart}, J.~M. and {Ho}, P.~T.~P. and {Patel}, N. and {G{\'o}mez}, Y.},
        title = "{Millimeter and Submillimeter High Angular Resolution Interferometric Observations: Dust in the Heart of IRAS 18162-2048}",
      journal = {\aj},
         year = 2011,
        month = mar,
       volume = {141},
       number = {3},
          eid = {72},
        pages = {72},
          doi = {10.1088/0004-6256/141/3/72},
archivePrefix = {arXiv},
       eprint = {1011.2237},
 primaryClass = {astro-ph.SR},
       adsurl = {https://ui.adsabs.harvard.edu/abs/2011AJ....141...72F}
}

@ARTICLE{2018Girart,
       author = {{Girart}, J.~M. and {Fern{\'a}ndez-L{\'o}pez}, M. and {Li}, Z. -Y. and {Yang}, H. and {Estalella}, R. and {Anglada}, G. and {{\'A}{\~n}ez-L{\'o}pez}, N. and {Busquet}, G. and {Carrasco-Gonz{\'a}lez}, C. and {Curiel}, S. and {Galvan-Madrid}, R. and {G{\'o}mez}, J.~F. and {de Gregorio-Monsalvo}, I. and {Jim{\'e}nez-Serra}, I. and {Krasnopolsky}, R. and {Mart{\'\i}}, J. and {Osorio}, M. and {Padovani}, M. and {Rao}, R. and {Rodr{\'\i}guez}, L.~F. and {Torrelles}, J.~M.},
        title = "{Resolving the Polarized Dust Emission of the Disk around the Massive Star Powering the HH 80-81 Radio Jet}",
      journal = {\apjl},
         year = 2018,
        month = apr,
       volume = {856},
       number = {2},
          eid = {L27},
        pages = {L27},
          doi = {10.3847/2041-8213/aab76b},
archivePrefix = {arXiv},
       eprint = {1803.06165},
 primaryClass = {astro-ph.SR},
       adsurl = {https://ui.adsabs.harvard.edu/abs/2018ApJ...856L..27G}
}

@ARTICLE{2009Mendoza,
       author = {{Mendoza}, S. and {Tejeda}, E. and {Nagel}, E.},
        title = "{Analytic solutions to the accretion of a rotating finite cloud towards a central object - I. Newtonian approach}",
      journal = {\mnras},
         year = 2009,
        month = feb,
       volume = {393},
       number = {2},
        pages = {579-586},
          doi = {10.1111/j.1365-2966.2008.14210.x},
archivePrefix = {arXiv},
       eprint = {0803.1020},
 primaryClass = {astro-ph},
       adsurl = {https://ui.adsabs.harvard.edu/abs/2009MNRAS.393..579M}
}

@ARTICLE{1976Ulrich,
       author = {{Ulrich}, R.~K.},
        title = "{An infall model for the T Tauri phenomenon.}",
      journal = {\apj},
         year = 1976,
        month = dec,
       volume = {210},
        pages = {377-391},
          doi = {10.1086/154840},
       adsurl = {https://ui.adsabs.harvard.edu/abs/1976ApJ...210..377U}
}

@ARTICLE{Busquet2019,
       author = {{Busquet}, G. and {Girart}, J.~M. and {Estalella}, R. and {Fern{\'a}ndez-L{\'o}pez}, M. and {Galv{\'a}n-Madrid}, R. and {Anglada}, G. and {Carrasco-Gonz{\'a}lez}, C. and {A{\~n}ez-L{\'o}pez}, N. and {Curiel}, S. and {Osorio}, M. and {Rodr{\'\i}guez}, L.~F. and {Torrelles}, J.~M.},
        title = "{Unveiling a cluster of protostellar disks around the massive protostar GGD 27 MM1}",
      journal = {\aap},
         year = 2019,
        month = mar,
       volume = {623},
          eid = {L8},
        pages = {L8},
          doi = {10.1051/0004-6361/201833687},
archivePrefix = {arXiv},
       eprint = {1902.07581},
 primaryClass = {astro-ph.SR},
       adsurl = {https://ui.adsabs.harvard.edu/abs/2019A&A...623L...8B}
}

@ARTICLE{Girart2018,
       author = {{Girart}, J.~M. and {Fern{\'a}ndez-L{\'o}pez}, M. and {Li}, Z. -Y. and {Yang}, H. and {Estalella}, R. and {Anglada}, G. and {{\'A}{\~n}ez-L{\'o}pez}, N. and {Busquet}, G. and {Carrasco-Gonz{\'a}lez}, C. and {Curiel}, S. and {Galvan-Madrid}, R. and {G{\'o}mez}, J.~F. and {de Gregorio-Monsalvo}, I. and {Jim{\'e}nez-Serra}, I. and {Krasnopolsky}, R. and {Mart{\'\i}}, J. and {Osorio}, M. and {Padovani}, M. and {Rao}, R. and {Rodr{\'\i}guez}, L.~F. and {Torrelles}, J.~M.},
        title = "{Resolving the Polarized Dust Emission of the Disk around the Massive Star Powering the HH 80-81 Radio Jet}",
      journal = {\apjl},
         year = 2018,
        month = apr,
       volume = {856},
       number = {2},
          eid = {L27},
        pages = {L27},
          doi = {10.3847/2041-8213/aab76b},
archivePrefix = {arXiv},
       eprint = {1803.06165},
 primaryClass = {astro-ph.SR},
       adsurl = {https://ui.adsabs.harvard.edu/abs/2018ApJ...856L..27G}
}

@ARTICLE{Qiu2009,
       author = {{Qiu}, Keping and {Zhang}, Qizhou},
        title = "{Discovery of Extremely High Velocity ``Molecular Bullets'' in the HH 80-81 High-Mass Star-Forming Region}",
      journal = {\apjl},
         year = 2009,
        month = sep,
       volume = {702},
       number = {1},
        pages = {L66-L71},
          doi = {10.1088/0004-637X/702/1/L66},
archivePrefix = {arXiv},
       eprint = {0907.5040},
 primaryClass = {astro-ph.GA},
       adsurl = {https://ui.adsabs.harvard.edu/abs/2009ApJ...702L..66Q}
}

@ARTICLE{Gomez1995,
       author = {{Gomez}, Y. and {Rodriguez}, L.~F. and {Marti}, J.},
        title = "{Thermal Jets and H 2O Masers: The Case of HH 80-81}",
      journal = {\apj},
         year = 1995,
        month = nov,
       volume = {453},
        pages = {268},
          doi = {10.1086/176386},
       adsurl = {https://ui.adsabs.harvard.edu/abs/1995ApJ...453..268G}
}

@ARTICLE{Canto2009,
       author = {{Cant{\'o}}, J. and {Curiel}, S. and {Mart{\'\i}nez-G{\'o}mez}, E.},
        title = "{A simple algorithm for optimization and model fitting: AGA (asexual genetic algorithm)}",
      journal = {\aap},
         year = 2009,
        month = jul,
       volume = {501},
       number = {3},
        pages = {1259-1268},
          doi = {10.1051/0004-6361/200911740},
archivePrefix = {arXiv},
       eprint = {0905.3712},
 primaryClass = {astro-ph.IM},
       adsurl = {https://ui.adsabs.harvard.edu/abs/2009A&A...501.1259C}
}

@ARTICLE{Zhao2024,
       author = {{Zhao}, X. and {Tang}, X.~D. and {Henkel}, C. and {Gong}, Y. and {Lin}, Y. and {Li}, D.~L. and {He}, Y.~X. and {Ao}, Y.~P. and {Lu}, X. and {Liu}, T. and {Sun}, Y. and {Wang}, K. and {Chen}, X.~P. and {Esimbek}, J. and {Zhou}, J.~J. and {Wu}, J.~W. and {Qiu}, J.~J. and {Zheng}, X.~W. and {Li}, J.~S. and {Luo}, C.~S. and {Zhao}, Q.},
        title = "{Kinetic temperature of massive star-forming molecular clumps measured with formaldehyde. V. The massive filament DR21}",
      journal = {\aap},
         year = 2024,
        month = jul,
       volume = {687},
          eid = {A207},
        pages = {A207},
          doi = {10.1051/0004-6361/202449352},
archivePrefix = {arXiv},
       eprint = {2405.18767},
 primaryClass = {astro-ph.GA},
       adsurl = {https://ui.adsabs.harvard.edu/abs/2024A&A...687A.207Z}
}

@ARTICLE{Tang2018,
       author = {{Tang}, X.~D. and {Henkel}, C. and {Wyrowski}, F. and {Giannetti}, A. and {Menten}, K.~M. and {Csengeri}, T. and {Leurini}, S. and {Urquhart}, J.~S. and {K{\"o}nig}, C. and {G{\"u}sten}, R. and {Lin}, Y.~X. and {Zheng}, X.~W. and {Esimbek}, J. and {Zhou}, J.~J.},
        title = "{ATLASGAL-selected massive clumps in the inner Galaxy. VI. Kinetic temperature and spatial density measured with formaldehyde}",
      journal = {\aap},
         year = 2018,
        month = mar,
       volume = {611},
          eid = {A6},
        pages = {A6},
          doi = {10.1051/0004-6361/201732168},
archivePrefix = {arXiv},
       eprint = {1711.10012},
 primaryClass = {astro-ph.GA},
       adsurl = {https://ui.adsabs.harvard.edu/abs/2018A&A...611A...6T}
}

@ARTICLE{2015Mangum,
       author = {{Mangum}, Jeffrey G. and {Shirley}, Yancy L.},
        title = "{How to Calculate Molecular Column Density}",
      journal = {\pasp},
         year = 2015,
        month = mar,
       volume = {127},
       number = {949},
        pages = {266},
          doi = {10.1086/680323},
archivePrefix = {arXiv},
       eprint = {1501.01703},
 primaryClass = {astro-ph.IM},
       adsurl = {https://ui.adsabs.harvard.edu/abs/2015PASP..127..266M}
}

@ARTICLE{Masque2012,
       author = {{Masqu{\'e}}, Josep M. and {Girart}, Josep M. and {Estalella}, Robert and {Rodr{\'\i}guez}, Luis F. and {Beltr{\'a}n}, Maria T.},
        title = "{Centimeter Continuum Observations of the Northern Head of the HH 80/81/80N Jet: Revising the Actual Dimensions of a Parsec-scale Jet}",
      journal = {\apjl},
         year = 2012,
        month = oct,
       volume = {758},
       number = {1},
          eid = {L10},
        pages = {L10},
          doi = {10.1088/2041-8205/758/1/L10},
archivePrefix = {arXiv},
       eprint = {1209.1254},
 primaryClass = {astro-ph.SR},
       adsurl = {https://ui.adsabs.harvard.edu/abs/2012ApJ...758L..10M}
}

@ARTICLE{Masque2015,
       author = {{Masqu{\'e}}, Josep M. and {Rodr{\'\i}guez}, Luis F. and {Araudo}, Anabella and {Estalella}, Robert and {Carrasco-Gonz{\'a}lez}, Carlos and {Anglada}, Guillem and {Girart}, Josep M. and {Osorio}, Mayra},
        title = "{Proper Motions of the Outer Knots of the HH 80/81/80N Radio-jet}",
      journal = {\apj},
         year = 2015,
        month = nov,
       volume = {814},
       number = {1},
          eid = {44},
        pages = {44},
          doi = {10.1088/0004-637X/814/1/44},
archivePrefix = {arXiv},
       eprint = {1510.01769},
 primaryClass = {astro-ph.SR},
       adsurl = {https://ui.adsabs.harvard.edu/abs/2015ApJ...814...44M}
}

@ARTICLE{Anez2020,
       author = {{A{\~n}ez-L{\'o}pez}, N. and {Osorio}, M. and {Busquet}, G. and {Girart}, J.~M. and {Mac{\'\i}as}, E. and {Carrasco-Gonz{\'a}lez}, C. and {Curiel}, S. and {Estalella}, R. and {Fern{\'a}ndez-L{\'o}pez}, M. and {Galv{\'a}n-Madrid}, R. and {Kwon}, J. and {Torrelles}, J.~M.},
        title = "{Modeling the Accretion Disk around the High-mass Protostar GGD 27-MM1}",
      journal = {\apj},
         year = 2020,
        month = jan,
       volume = {888},
       number = {1},
          eid = {41},
        pages = {41},
          doi = {10.3847/1538-4357/ab5dbc},
archivePrefix = {arXiv},
       eprint = {1911.12398},
 primaryClass = {astro-ph.SR},
       adsurl = {https://ui.adsabs.harvard.edu/abs/2020ApJ...888...41A}
}

@ARTICLE{Gieser2021,
       author = {{Gieser}, C. and {Beuther}, H. and {Semenov}, D. and {Ahmadi}, A. and {Suri}, S. and {M{\"o}ller}, T. and {Beltr{\'a}n}, M.~T. and {Klaassen}, P. and {Zhang}, Q. and {Urquhart}, J.~S. and {Henning}, Th. and {Feng}, S. and {Galv{\'a}n-Madrid}, R. and {de Souza Magalh{\~a}es}, V. and {Moscadelli}, L. and {Longmore}, S. and {Leurini}, S. and {Kuiper}, R. and {Peters}, T. and {Menten}, K.~M. and {Csengeri}, T. and {Fuller}, G. and {Wyrowski}, F. and {Lumsden}, S. and {S{\'a}nchez-Monge}, {\'A}. and {Maud}, L. and {Linz}, H. and {Palau}, A. and {Schilke}, P. and {Pety}, J. and {Pudritz}, R. and {Winters}, J.~M. and {Pi{\'e}tu}, V.},
        title = "{Physical and chemical structure of high-mass star-forming regions. Unraveling chemical complexity with CORE: the NOEMA large program}",
      journal = {\aap},
         year = 2021,
        month = apr,
       volume = {648},
          eid = {A66},
        pages = {A66},
          doi = {10.1051/0004-6361/202039670},
archivePrefix = {arXiv},
       eprint = {2102.11676},
 primaryClass = {astro-ph.GA},
       adsurl = {https://ui.adsabs.harvard.edu/abs/2021A&A...648A..66G}
}

@ARTICLE{Gerner2014,
       author = {{Gerner}, T. and {Beuther}, H. and {Semenov}, D. and {Linz}, H. and {Vasyunina}, T. and {Bihr}, S. and {Shirley}, Y.~L. and {Henning}, Th.},
        title = "{Chemical evolution in the early phases of massive star formation. I}",
      journal = {\aap},
         year = 2014,
        month = mar,
       volume = {563},
          eid = {A97},
        pages = {A97},
          doi = {10.1051/0004-6361/201322541},
archivePrefix = {arXiv},
       eprint = {1401.6382},
 primaryClass = {astro-ph.SR},
       adsurl = {https://ui.adsabs.harvard.edu/abs/2014A&A...563A..97G}
}

\begin{appendix}
\onecolumn
\section{Channel maps}
\label{app:channel}

\begin{figure*}[ht!]
\centering
\includegraphics[width=\linewidth]{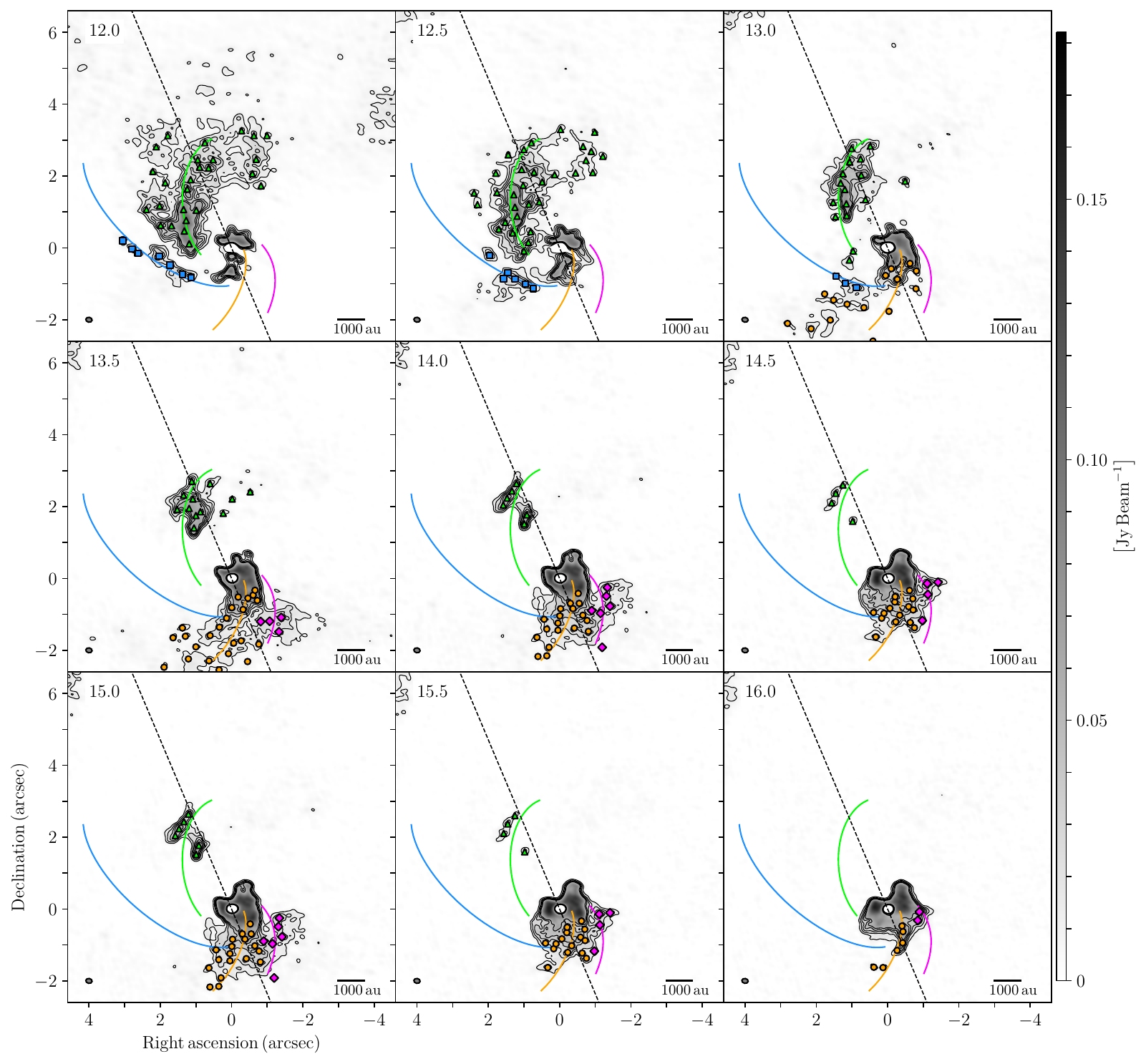}
\caption{Channel maps of the H$_2$CO molecular line emission of the MM1 protostellar system. The channel velocity in \kms is indicated in the top-left corner. The magenta, orange, blue, and green lines are the best-fit models for the four streamers St1, St2, St3, St4, respectively. The diamonds, circles, squares and triangles correspond to the positions of the condensations. The synthesized beam is shown in the bottom-left corner of all panels. The contours start at 5$\sigma$ and increase in steps of 5$\sigma$ up to 25$\sigma$, where $\sigma = 2.1$ mJy beam$^{-1}$.}
\label{fig:channelsmaps}
\end{figure*}

Figure \ref{fig:channelsmaps} shows the redshifted emission of the H$_2$CO molecular line from the velocity cube, with a channel width of 0.5 \kms. The four fitted-streamers are depicted in all channels as purple, orange, blue, and green lines, corresponding to St1, St2, St3, and St4, respectively. The positions of the condensations, along with their corresponding \com{line-of-sight} velocities, are also plotted.

\end{appendix}

\end{document}